\documentclass[epj]{svjour}
\usepackage{hyperref}
\usepackage{graphics,graphicx,epsfig,amssymb,amsmath,cite}
\usepackage{slashed}

\newcommand{\madanalysis}{{\sc MadAnalysis~5}}

\newcommand{\delphes}{{\sc Delphes}}

\newcommand{\eg}{{\it e.g.}}
\newcommand{\ie}{{\it i.e.}}

\def\ma{\madanalysis}

\begin{document}
\title{Towards a public analysis database for LHC new physics searches using {\sc MadAnalysis~5}}
\author{
B.~Dumont\inst{1} \and
B.~Fuks\inst{2,3} \and
S.~Kraml\inst{1} (eds.) \and 
S.~Bein\inst{4} \and 
G.~Chalons \inst{1} \and
E.~Conte\inst{5} \and
S.~Kulkarni\inst{1} \and
D.~Sengupta\inst{1} \and
C.~Wymant\inst{6}\thanks{\emph{Present address:}    Department of Infectious Disease Epidemiology,
   Imperial College London,
   St Mary's Campus,
   Norfolk Place
   London, W2 1PG, UK.}
%
% \thanks is optional - remove next line if not needed
%\thanks{\emph{Present address:} Insert the address here if needed}%
%
}                     % end of author block
\institute{
  Laboratoire de Physique Subatomique et de Cosmologie, %LPSC 
  Universit\'e Grenoble-Alpes, CNRS/IN2P3, 53 Avenue des Martyrs, F-38026 Grenoble Cedex, France 
     \and  
CERN, PH-TH, CH-1211 Geneva 23, Switzerland 
     \and
Institut Pluridisciplinaire Hubert Curien/D\'epartement Recherches Subatomiques, Universit\'e de Strasbourg/CNRS-IN2P3, 23 Rue du Loess, F-67037 Strasbourg, France
     \and
 Department of Physics, Florida State University, Tallahassee, Florida 32306, USA 
     \and
Groupe de Recherche de Physique des Hautes Energies (GRPHE), Universit\'{e} de Haute-Alsace, IUT Colmar, 34 rue du Grillenbreit BP 50568, 68008 Colmar Cedex, France 
     \and
LAPTH, 9 Chemin de Bellevue, B.P. 110, Annecy-le-Vieux 74951, France
}

\date{Received: date / Revised version: date}
% The correct dates will be entered by Springer

\abstract{
We present the implementation, in the {\sc MadAnalysis}~5 framework, of several ATLAS and CMS searches for supersymmetry in data recorded during the first run of the LHC. We provide extensive details on the validation of our implementations and propose to create a public analysis database within this framework.
 % end of PACS codes
} %end of abstract
\maketitle

%--- preprint numbers ----
%\vspace*{-11.5cm} \noindent 
%\small{CERN-PH-TH-2014-109, LAPTH-048/14}\\
%\vspace*{9cm}

%%%%%%%%%%%%%%%%%%%%%%%%%%%%%%%%%%%%%%%%%%%%%%%%%%%%%%%%%

\section{Introduction}  \label{intro}

%%%%%%%%%%%%%%%%%%%%%%%%%%%%%%%%%%%%%%%%%%%%%%%%%%%%%%%%%

The LHC was designed as a machine of discovery.
It was built to explore the TeV energy scale, in order to unravel
the mechanism of electroweak symmetry breaking and shed light on new
physics beyond the Standard Model (SM).   
The recent discovery~\cite{Aad:2012tfa,Chatrchyan:2012ufa} of a new particle with mass of 125~GeV 
and properties consistent with the SM Higgs boson is a first triumph for the LHC physics program 
and has profound implications for our understanding of the universe.  
We are, however, still left with many fundamental
questions open, and to address them it is imperative that the search
for new physics continue, at the LHC and elsewhere. 

During Run~I of the LHC at center-of-mass energies of $\sqrt{s} = 7$ and 8~TeV, 
the ATLAS and CMS collaborations have carried out an extensive program searching for 
new physics in many different channels~\cite{atlas:susytwiki,atlas:exoticstwiki,cms:susytwiki,cms:exoticstwiki}. 
Since no signal was found, the experimental collaborations interpreted their results  
setting limits in terms of popular models, such as the CMSSM 
(Constrained Minimal Supersymmetric Standard Model, see \eg~\cite{AbdusSalam:2011fc}), 
or in terms of so-called Simplified Model Spectra (SMS).\footnote{Simplified Models are effective-Lagrangian descriptions involving  only a small number of new particles. They were designed as a useful tool for the characterization 
of new physics, see \eg\ \cite{Alwall:2008ag,Alves:2011wf}.}  
These searches will be pursued further at higher energies,  
with first results to be expected soon after the start of Run~II in 2015. 

There exist, however, many different  beyond-the-SM (BSM) theories, and each of them
comes with a large variety of concrete realizations. This leads to a multitude of possible scenarios, 
with complex interrelations between parameters and signatures. 
It is a challenge for the whole community to work out the implications of the LHC results 
in the contexts of all these different models, to derive the relevant limits, point out  
possible loopholes in the current searches, and help design the analyses for the 
next phase of LHC running at higher energy.

To this end, many groups have been developing private codes for the interpretation of the LHC results. 
Moreover, recently some public tools became available, which serve the whole community. 
For the interpretation in the context of Simplified Models, there are {\sc SModelS}~\cite{Kraml:2013mwa} 
and {\sc Fastlim}~\cite{Papucci:2014rja}. 
{\sc SModelS} takes the spectrum of any BSM scenario, decomposes it into SMS topologies, and compares it to the cross section upper limits from more than 50 ATLAS and CMS SMS results. 
{\sc Fastlim} reconstructs the visible cross sections from pre-calculated efficiency and cross section tables for simplified event topologies, currently taking into account 11~ATLAS analyses which mainly focus on searches for supersymmetric partners of the top and bottom quarks (stops and sbottoms, respectively).   
For confronting simulated events of any model to LHC results, there is {\sc CheckMATE}~\cite{Drees:2013wra}. 
This program currently has 8 ATLAS and 1 CMS supersymmetry (SUSY) analyses implemented, 
which it re-interprets based on fast simulation. 
Another tool, {\sc XQCAT}~\cite{Barducci:2014ila}, is designed specifically for testing scenarios with 
heavy extra quarks, based on a CMS search for top partners as well as two SUSY searches.
Finally, {\sc ATOM}~\cite{ATOM} is being developed for calculating the efficiencies of signal and control 
regions of various LHC searches  based on the {\sc Rivet}~\cite{Buckley:2010ar} toolkit.\footnote{We note that {\sc Rivet}~\cite{Buckley:2010ar} itself is designed for unfolded data. Unfolding works very well for SM measurements, and consequently there are many SM analyses from the LHC available on {\sc Rivet}. For BSM searches with large missing energy however (the typical SUSY case) unfolding is still an open issue.} 
%%% new %%%
In addition to these stand-alone tools, developed mostly by theorists, one should note 
the {\sc RECAST} framework \cite{Cranmer:2010hk}, which aims at providing a generic platform for 
requests of re-interpretation of existing analyses; in this case the re-interpretation would be done by the experimental collaboration itself, using the official full simulation software. 
%%%
  
In this paper, we follow a complementary approach. We present the implementation  
of several  ATLAS and CMS SUSY analyses in \ma~\cite{Conte:2012fm,Conte:2014zja}, 
with simulation of detector effects based on \delphes~3~\cite{deFavereau:2013fsa}, 
and propose to create a public analysis database (PAD) within this framework. 
\ma\ offers a user-friendly platform for collider phenomenology, 
and the PAD we propose will be easily accessible to and extendible by the whole community. 
%%% new %%%
Our proposal is complementary to the existing tools mentioned above in that 
{\it i)} it is based on event simulation, thus avoiding the shortcomings of the Simplified Models approach; 
{\it ii)} the output is the number of events in the different experimental regions of an analysis , which can then be statistically interpreted by the user for a variety of tasks, including limit setting or developing efficiency maps for Simplified Models; and {\it iii)}~it is a completely Open Source initiative. 
%%% 

In BSM searches, sets of selection criteria are designed in order to maximize the sensitivity to expected signals of new physics. These define so-called signal and control regions, described in the experimental publications. 
For interpreting a search in the context of a given new physics model, one has to implement these selection criteria together with a description of the detector performance (emulating the various object definitions and efficiencies) in a simulation tool. 
Based on simulated event samples for the model being tested,  the expected number of signal events  in the various signal regions (SRs) can then be computed and compared to the number of observed events and the number of expected SM background events, which are reported in the experimental publication.  

Non-collaboration members however do not have access to the experimental data, nor the Monte Carlo (MC) event set 
simulated with an official collaboration detector simulation. This renders the implementation and validation 
of ATLAS and CMS analyses for re-interpretation of the experimental results in general contexts 
a tedious task, even more so as the information given in the experimental papers is often incomplete (we will comment more on this in Section~3).  
We therefore think that a common platform for collecting object definitions, cuts, and all other information 
necessary to reproduce or use the results of the analyses will be of great value to the high-energy physics community.  
Moreover, as our project follows an Open Access and Open Data approach, we hope that it will benefit the 
scientific communication and in particular motivate ATLAS and CMS to provide more information 
on their analyses, in line with the Les Houches Recommendations~\cite{oai:arXiv.org:1203.2489}. 

The rest of the paper is organized as follows. 
In Section~\ref{sec:MA5new}, we briefly recall some new features in \ma, which are pertinent for implementing LHC analyses, and describe the modifications to the \delphes~3 detector simulation which we adopted for this project. 
In Section~\ref{sec:implemended}, we present some ATLAS and CMS analyses which we implemented in the \ma\ framework and report in detail on their validation. 
The relevant C++ codes are all publicly available and may thus constitute the foundation of the PAD. 
A module for a simplified statistical interpretation of the simulated signals is presented in Section~\ref{sec:CLprocedure}.  
%%% new %%% 
In Section~\ref{sec:guidelines} we provide some guidelines, on the one hand for the experimental collaborations regarding what material is needed for a reliable implementation and validation of an analysis, on the other hand for potential contributors to the framework as to how to validate a new analysis implementation.  
%%%
Section~\ref{sec:conclusions} contains our conclusions.

%%%%%%%%%%%%%%%%%%%%%%%%%%%%%%%%%%%%%%%%%%%%%%%%%%%%%%%%%

\section{New developments in \ma\ and {\sc Delphes~3}}\label{sec:MA5new}

%%%%%%%%%%%%%%%%%%%%%%%%%%%%%%%%%%%%%%%%%%%%%%%%%%%%%%%%%

%----------------------------------------------------------------------------------------------------------------------------------------------------------
\subsection{Dealing with multiple signal regions in \ma} \label{sec:MA5SRM}
%----------------------------------------------------------------------------------------------------------------------------------------------------------

In most experimental analyses performed at the LHC, and in particular
the searches considered in this work, a branching set of
selection criteria (``cuts'') is used to define several
different sub-analyses (``regions'') within the same analysis.
In conventional coding frameworks, multiple regions are implemented with a nesting
of conditions checking these cuts, which grows exponentially more complicated
with the number of cuts. The scope of this project has therefore motivated us to
extend the \ma\ package to facilitate the handling of analyses with multiple regions,
as first presented in~\cite{Brooijmans:2014eja} and described in detail
in~\cite{Conte:2014zja}.

From version 1.1.10 onwards, the implementation of an analysis in the \ma\ framework
consists of implementing three basic functions:
\begin{itemize}
  \item \texttt{Initialize}, dedicated to the initialization of the signal regions,
    histograms, cuts and any user-defined variables;
  \item \texttt{Execute}, containing the analysis cuts and weights applied to each event; and
  \item \texttt{Finalize}, controlling the production of the results of the analysis, \ie,
    histograms and cut-flow charts.
\end{itemize}
The new functionalities of \ma\ for implementing LHC analyses
are described in detail in the new manual of its expert mode~\cite{Conte:2014zja}.
To illustrate
the handling of multiple regions, we present a few
snippets of our implementation \cite{ma5code:cms-sus-13-011}
of the CMS search for stops in final states with one lepton~\cite{Chatrchyan:2013xna}
(see Section \ref{sec:cms-sus-13-011}).
This search comprises 16 SRs, all of which must be declared in the
\texttt{Initialize} function.
This is done through the \texttt{AddRegionSelection} method
of the analysis manager class, of which \texttt{Manager()} is an instance provided
by default with each analysis. It takes as its argument
a string uniquely defining the SR under consideration.
For instance, two of the 16 SRs of the CMS analysis are declared as
\begin{verbatim}
 Manager()->AddRegionSelection(
   "Stop->t+neutralino,LowDeltaM,MET>150");
 Manager()->AddRegionSelection(
   "Stop->t+neutralino,LowDeltaM,MET>200");
\end{verbatim}

The \texttt{I\-ni\-ti\-a\-li\-ze}
function
should also contain the declaration of selection cuts. This
is handled by the \texttt{AddCut} method of
the analysis manager class. If a cut is common to all SRs, the
\texttt{AddCut} method takes as a single argument a string that uniquely identifies the cut.
An example of the declaration of two common cuts is
\begin{verbatim}
 Manager()->AddCut("1+ candidate lepton");
 Manager()->AddCut("1 signal lepton");
\end{verbatim}
If a cut is not common to all regions,
the \texttt{AddCut} method requires a second argument, either a
string or an array of strings, consisting of the names of all the regions to which
the cut applies. For example, an $E_T^{\rm miss}>150$~GeV cut that applies to four SRs
could be declared as
\begin{verbatim}
 string SRForMet150Cut[] = {
   "Stop->b+chargino,LowDeltaM,MET>150",
   "Stop->b+chargino,HighDeltaM,MET>150",
   "Stop->t+neutralino,LowDeltaM,MET>150",
   "Stop->t+neutralino,HighDeltaM,MET>150"};
 Manager()->AddCut("MET>150GeV",SRForMet150Cut);
\end{verbatim}

Histograms are initialized in a similar fashion using the \texttt{AddHisto} method
of the manager class. A string argument is hence required
to act as a unique identifier for the histogram, provided together with its number
of bins and bounds. A further optional argument consisting
of a string or array of strings can then be used to associate it with specific
regions. The exact syntax can be found in the manual~\cite{Conte:2014zja}.

Most of the logic of the analysis is implemented in the \texttt{Execute} function.
This relies both on standard methods to declare particle objects and to compute
the observables of interest for event samples including detector
simulation~\cite{Conte:2012fm} and on the new manner in which cuts are
applied and histograms filled via the analysis manager class~\cite{Conte:2014zja}.
In particular, we emphasize the existence of a new \texttt{isolCones} method
of the \texttt{RecLeptonFormat} class
for testing the isolation
of the leptons. This returns a vector of
\texttt{IsolationConeType} objects describing the transverse activity
in a cone of radius $\Delta R$ centered on the lepton and whose properties
are the following:
\begin{itemize}
 \item \texttt{deltaR()}: returns the size of the cone;
 \item \texttt{ntracks()}: returns the number of tracks present in the cone;
 \item \texttt{sumPT()}: returns the scalar sum of the transverse momenta of all tracks lying
   in the cone;
 \item \texttt{sumET()}: returns the scalar sum of the transverse energy deposits
   in the cone.
\end{itemize}
In general, experimental analyses only consider tracks with a transverse momentum larger
than a given threshold. It should be noted that \ma\ does not control this last
functionality so that the threshold must be specified at the level of the
detector simulator.
All these features
should be used together with the modifications of \delphes~3
described in the next subsection.

Below we provide a couple of examples for applying cuts and filling
histograms. After having declared and filled two vectors,
\texttt{Si\-gnal\-E\-lec\-trons} and \texttt{SignalMuons}, with objects satisfying the signal
lepton definitions used in the CMS-SUS-13-011 analysis,
we
require exactly one signal lepton with the following selection cut:
\begin{verbatim}
if( !Manager()->ApplyCut( 
 (SignalElectrons.size()+SignalMuons.size())>0,
      "1+ candidate lepton") ) return true;
\end{verbatim}
The \texttt{if(...)} syntax guarantees that a given event is discarded 
as soon as all regions fail the cuts applied so far.
Histogramming is as easy as applying a cut. For example, as we are interested in
the transverse-momentum distribution of the leading lepton, our code contains
\begin{verbatim}
  Manager()->FillHisto("pT(l)",
       SignalLeptons[0]->pt());
\end{verbatim}
This results in the filling of a histogram, previously declared with
the name \texttt{"pT(l)"} in the \texttt{Initialize} method, but only
when all cuts applied to the relevant regions are satisfied.

Finally, event weights often need to be applied at the analysis level to correct
for the efficiency with which physical objects, such as electrons or jets, are
identified or likely to trigger the event. In \ma, the weight of an event can
easily be modified, if necessary, by using the {\tt SetCurrentEventWeight} method of the
manager class.

After the execution of the program, a set of {\sc Saf} files (an {\sc Xml}-inspired
format used by \ma) is created.
These files are organized in an automatically generated output directory with
the same name as the input file (containing the path(s) to the event file(s) to consider),
chosen to be \texttt{input.txt} for the sake of the example.
At the root of this output directory, one finds a file named in our case
\texttt{input.txt.saf} with general information on
the analyzed events, such as the associated cross section, the number of events,
\textit{etc}. It comes together with a series of subdirectories named according
to the different analyses that have been executed. In the case of an analysis
denoted by \texttt{cms\_sus\_13\_011}, the corresponding subdirectory will contain:
\begin{itemize}
  \item a {\sc Saf} file \texttt{cms\_sus\_13\_011.saf} listing the names of all the
    implemented SRs;
  \item a subdirectory \texttt{Histograms} with a {\sc Saf} file
    \texttt{histos.saf} describing all the histograms that have been
    implemented; and
  \item a subdirectory \texttt{Cutflows} with a series of {\sc Saf} files (named
    according to the definition of the SRs) containing the cut flow tables of all declared SRs.
\end{itemize}
The structure of the various {\sc Saf} files is detailed in \cite{Conte:2014zja}.

%----------------------------------------------------------------------------------------------------------------------------------------------------------
\subsection{The `MA5tune' of {\sc Delphes~3}} \label{sec:moddelphes3}
%----------------------------------------------------------------------------------------------------------------------------------------------------------

\delphes~\cite{deFavereau:2013fsa} is a {\sc C++} framework dedicated to
the simulation of a generic detector such as those used in collider experiments.
Contrary to full detector simulation software, \delphes\ does not simulate
the particle-matter interactions, but uses instead
a parameterization of the detector response and reconstructs the main
physics objects considered in the analyses.
This simplified picture results in much faster simulations, while the accuracy level is 
maintained suitable for realistic phenomenological investigations. 
From the computing side, \delphes\ is a modular framework where developers can
both add their own contributions and tune the default parameterization according
to their needs. This modularity is based on a division of the simulation process
into modules inspired by the \texttt{TTask} \textsc{Root} class, and the addition and
removal of new elements are easily achievable through a \textsc{Tcl}
configuration file. Similarly, the content of the output
\textsc{Root} files can be configured at will.

In order to properly recast ATLAS and CMS analyses, a tuning of the version 3 of
\delphes\ has been performed.
In the original version of \delphes, an isolation criterion is applied
to both leptons and photons, and only particles satisfying this requirement
are stored in the output files. We have designed a new \delphes\ module
named \texttt{CalculationIsolation} that allows one to move the isolation
requirements in the analysis selection. 
This module computes several variables useful for the implementation of
isolation cuts. Defining cone sizes of $\Delta R=0.2, 0.3, 0.4$ and $0.5$,
the number of tracks with a transverse momentum larger than a given
threshold, the scalar sum of the transverse momenta of these
tracks and the scalar sum of the calorimetric transverse energy deposits
lying in the cones are evaluated and saved. In addition, the 
default module of \delphes\ dedicated to the
filtering of non-i\-so\-la\-ted lepton and photon candidates
is switched off so that all candidates are kept in the output \textsc{Root}
files. For consistency reasons, the \delphes\ module \texttt{U\-ni\-que\-Ob\-ject\-Find\-er}
giving a unique identification to all reconstructed objects is bypassed.
Isolation selection cuts can then be performed at the analysis
level by means of the \texttt{isolCones} method of the
\texttt{RecLeptonFormat} class of \ma, described in the previous
subsection and in~\cite{Conte:2014zja}.

Adding the isolation information to the output format yields an
increase of the size of the output files. A cleaning of all collections is
therefore in order to reduce the file sizes. First, collections such as
calorimeter towers and particle-flow objects are not stored. Next,
the (heavy) collection of all particles that have been generated at the
different level of the simulation chain (hard scattering process, parton
showering and hadronization) is pruned. Only particles produced
at the hard-scattering process level, as well as final-state leptons
and $b$ quarks present after parton showering, are stored. In addition,
the relations between generated and reconstructed leptons have been
retained, together with information on the origin (the mother
particle) of each lepton. All these changes result in a reduction
of the size of the produced \textsc{Root} files by about a factor of ten when
compared to the files produced with the original configuration of \delphes.

This tailored version of \delphes~3, which we internally call
{\sc Delphes-MA5tune} to avoid confusion with the original version,
can conveniently be installed locally from the \ma\ interpreter by
typing in the command
\begin{verbatim}
  install delphesMA5tune
\end{verbatim}
Even if \delphes~3 is already installed on a given system, one will
need this modified `MA5tune' version of the program in order to run the
\ma\ analyses that we are presenting in this paper. 
Note however that for the moment  \ma\ is  not able to run with 
both \delphes\ and {\sc Delphes-MA5tune}  installed in parallel.  
This means that  
the user must take care 
that only the directory \texttt{tools/delphesMA5tune} (but not the directory \texttt{tools/delphes}) 
be available in his/her local installation of \ma.

In order to process an (hadronized) event sample with
the `MA5tune' of \delphes, it is sufficient to start \ma\
in the reconstructed mode, import the considered sample and type
\begin{verbatim}
  set main.fastsim.package = delphesMA5tune
  set main.fastsim.detector = cms
  submit
\end{verbatim}
where \texttt{cms} can be replaced by \texttt{atlas} according to the needs of
the user. Default detector parameters are employed and can be modified by the user, following
the guidelines displayed on the screen. The output \textsc{Root} file can then be
retrieved from the automatically generated working directory.

%%%%%%%%%%%%%%%%%%%%%%%%%%%%%%%%%%%%%%%%%%%%%%%%%%%%%%%%%

\section{Implemented Analyses and their Validation}\label{sec:implemended}

%%%%%%%%%%%%%%%%%%%%%%%%%%%%%%%%%%%%%%%%%%%%%%%%%%%%%%%%%

To start the analysis database, we have implemented and validated the following ATLAS and CMS SUSY 
searches at $\sqrt{s}=8$~TeV and an integrated luminosity of about 20~fb$^{-1}$: \\

\noindent {\bf ATLAS:}
\begin{itemize}
\item Search for stops and sbottoms in final states with no lepton and two $b$-jets~\cite{Aad:2013ija}: ATLAS-SUSY-2013-05;\\
\item Search for charginos, neutralinos and sleptons in final states with two leptons~\cite{Aad:2014vma}: ATLAS-SUSY-2013-11;\\
\end{itemize}

\noindent {\bf CMS:}
\begin{itemize}
\item Search for stops in the single-lepton final state~\cite{Chatrchyan:2013xna}: CMS-SUS-13-011;\\
\item Search for gluinos and squarks in events with three or more jets and $E_T^{\rm miss}$~\cite{Chatrchyan:2014lfa}: CMS-SUS-13-012;\\
\item Search for gluinos in opposite-sign dilepton events, large number of jets, $b$-jets and $E_T^{\rm miss}$~\cite{CMS-PAS-SUS-13-016}: CMS-SUS-13-016. \\
\end{itemize}

\noindent 
Several more analyses are currently being implemented and validated.

Below we give some details on these analyses, the level of documentation by the experimental collaboration, 
and the validation of our \ma\ implementations. We begin with the CMS stop search in the single-lepton channel, 
which also served as our template analysis for developing the extensions of \ma\ described briefly 
in Section~\ref{sec:MA5new} and in detail in \cite{Conte:2014zja}.  
The related recast code \cite{ma5code:cms-sus-13-011} contains extensive comments, 
which should allow the interested reader to easily use it as template for implementing a different analysis. 

A list of all available analyses (which will certainly evolve quickly), instructions on how to use them, 
as well as more detailed validation notes can be found on the \ma\ wiki page~\cite{ma5wiki}. 
The recast codes themselves are published via {\sc Inspire}~\cite{inspire}, in order to make them citable 
({\sc Inspire} assigns each submission a DOI~\cite{doi}) and 
to ensure that changes can be traced reliably through a rigorous versioning system. 

Before proceeding, some general comments are in order. 
Generally, we cannot reproduce cleaning cuts (for, \eg, cosmic rays and beam effects). 
Moreover, some basic jet quality criteria must be skipped as we do not have vertex information. 
This is, however, expected to have a small impact on signal events. 
In addition, event weights are typically applied by ATLAS and CMS to correct simulated events with respect to data. 
We take such event weights into account whenever they are available. 
Otherwise they are neglected and contribute to the overall systematic
uncertainty. We note that this uncertainty is expected to be larger
when testing signals that are very different from the ones used for
the validation, depending on the nature of the reconstructed objects
and on the kinematic configuration of the events. In such a case one should
interpret the result with care.

Finally, while the selection criteria that define the various SRs are usually clear and well documented, 
information on the preselection cuts is often missing. In particular, trigger efficiencies, information about isolation, 
efficiencies for leptons, and the order in which preselection cuts are applied is crucial for reliably reproducing an 
analysis, but this information is often incomplete in the experimental publications. 
We hope that this will improve over time and the necessary information will be given systematically 
either in the physics paper or in a performance note, 
as also advertised in \cite{oai:arXiv.org:1203.2489}.

%----------------------------------------------------------------------------------------------------------------------------------------------------------
\subsection{CMS-SUS-13-011: search for stops in the single-lepton final state} \label{sec:cms-sus-13-011}
%----------------------------------------------------------------------------------------------------------------------------------------------------------

The CMS search for stops in the single lepton and missing energy, $\ell + E^{\rm miss}_T$, final state with full luminosity at
$\sqrt{s} = 8$~TeV~\cite{Chatrchyan:2013xna} has been taken as a ``template analysis'' to develop a common language and framework for the analysis implementation. It also allowed us to test the new developments in \ma\ which were necessary for carrying out this project.

The analysis targets two possible decay modes of the stop: $\tilde{t} \to t \tilde{\chi}^{0}_1$ and 
$\tilde{t} \to b \tilde{\chi}^{+}_1$.  
Since the stops are pair-produced, their decays give rise to two $W$-bosons in each event, one of which is assumed to  decay leptonically, whilst the other one is assumed to decay hadronically. 
In the cut-based version of the analysis,\footnote{The search also contains an analysis based on multivariate analysis techniques (MVA); such analyses generically cannot be externally reproduced unless the final MVA is given. As this is not the case so far, we here only use the cut-based version of the analysis.} two sets of signal regions with different cuts, each  dedicated to one of the two decay modes, are defined. These two sets are further divided into ``low $\Delta M$'' and ``high $\Delta M$'' categories, targeting small and large mass differences with the lightest neutralino $\tilde\chi_1^0$, respectively. Finally, each of these four categories are further sub-divided using four different $E^{\rm miss}_T$ requirements. In total, 16 different, potentially overlapping SRs are defined. 
Two cuts are based on rather complex and specific kinematic variables designed to reduce the dilepton $t\bar{t}$ background: a $\chi^2$ resulting from the full reconstruction of the hadronic top and $M^W_{T2}$ -- a variant of the $m_{T2}$ observable. The implementation of the $\chi^2$ quantity in our code was straightforward thanks to the {\sc C++} {\sc Root} code provided on the CMS Twiki page.  
The $M_{T2}^W$ variable is calculated with the standard \ma\ method, see~\cite{Conte:2014zja}, 
according to the algorithm presented in~\cite{Bai:2012gs}. 

Overall, this analysis is very well documented. Some important pieces of information were however missing, in particular the detailed trigger efficiencies and the identifi\-cation-only efficiencies for electron and muons. These were provided by the CMS collaboration upon request and are now available on the analysis Twiki page~\cite{cms-sus-13-011-twiki} in the section ``Additional Material to aid the Phenomenology Community with Reinterpretations of these Results''. In addition,  the $b$-tagging efficiency as a function of $p_T$ is not given in the paper, but was taken from~\cite{Chatrchyan:2013fea}.
Another technical difficulty came from the isolation criteria. Indeed, the CMS analysis considers the sum of transverse momenta of so-called `Particle Flow' particles in a cone of given $\Delta R$. This is difficult to reproduce in our case. Instead, we only use tracks in the inner detector for the isolation. From the two benchmark points for which cut flows are available (see Table~\ref{tab:cms-13-011-cutflow}) we found that a weighting factor of $0.885$, applied on the events at the same time as the isolation, is sufficient to correct our track-only isolation. Therefore we incorporate this correction to our analysis code.

\begin{table}[!h]
\caption{Final number of events for $\tilde{t} \rightarrow b \tilde{\chi}^{\pm}_1$  in three SRs of the analysis
CMS-SUS-13-011. The benchmark points are given in the format 
$(m_{\tilde t},m_{\tilde\chi^0_1},x)$ in GeV, with $x$ setting the chargino mass according to 
$m_{\tilde{\chi}^+_1} = x \cdot m_{\tilde{t}} + (1 - x) m_{\tilde{\chi}^0_1}$.
\label{tab:NevStopbcharginoLowDeltaMMET2501}}
\begin{center}
  \begin{tabular}{l|c|c}
  benchmark point & CMS result & {\sc MA}5 result \\ 
  \hline\noalign{\smallskip}
        \multicolumn{3}{c}{$\tilde{t} \rightarrow b \tilde{\chi}^{\pm}_1, {\rm low\,} \Delta M, E^{\rm miss}_T > 150~{\rm GeV}$} \\ 
  $(250/50/0.5)$ & $157 \pm 9.9$ & $141.2$ \\ 
  $(250/50/0.75)$ & $399 \pm 18$ & $366.8$ \\ 
  \noalign{\smallskip}
  \hline\noalign{\smallskip}
        \multicolumn{3}{c}{$\tilde{t} \rightarrow b \tilde{\chi}^{\pm}_1, {\rm high\,} \Delta M, E^{\rm miss}_T > 150~{\rm GeV}$} \\ 
  $(450/50/0.25)$ & $23 \pm 2.3$ & $23.4$ \\ 
  \noalign{\smallskip} 
  \hline\noalign{\smallskip}
        \multicolumn{3}{c}{$\tilde{t} \rightarrow b \tilde{\chi}^{\pm}_1, {\rm high\,} \Delta M, E^{\rm miss}_T > 250~{\rm GeV}$} \\ 
  $(600/100/0.5)$ & $6.1 \pm 0.5$ & $5.4$ \\ 
  $(650/50/0.5)$ & $6.7 \pm 0.4$ & $5.8$ \\ 
  $(650/50/0.75)$ & $6.3 \pm 0.4$ & $5.7$ \\ 
  \noalign{\smallskip}\hline
\end{tabular}
\end{center}
\end{table}

\begin{table}[!h]
\caption{Final number of events for $\tilde{t} \rightarrow t \tilde{\chi}^0_1$ in two SRs of the analysis CMS-SUS-13-011. 
For each benchmark point, the first number indicates the stop mass, the second the LSP mass (in GeV).
\label{tab:NevStopTneutralinoLowDeltaMMET2502}}
\begin{center}
\begin{tabular}{c|c|c}
  benchmark point & CMS result & {\sc MA}5 result \\
  \hline\noalign{\smallskip}
        \multicolumn{3}{c}{$\tilde{t} \rightarrow t \tilde{\chi}^0_1, {\rm low\,} \Delta M, E^{\rm miss}_T > 150~{\rm GeV}$} \\ 
  $(250/50)$ & $108 \pm 3.7$ & $100.1$ \\ 
\noalign{\smallskip}
\hline\noalign{\smallskip}
        \multicolumn{3}{c}{$\tilde{t} \rightarrow t \tilde{\chi}^0_1, {\rm high\,} \Delta M, E^{\rm miss}_T > 300~{\rm GeV}$} \\ 
  $(650/50)$ & $3.7 \pm 0.1$ & $3.6$ \\ 
  \noalign{\smallskip} 
  \hline\noalign{\smallskip}
\end{tabular}
\end{center}
\end{table}

\begin{table*}[!t]
\caption{Summary of yields for the $\tilde{t} \rightarrow t \tilde{\chi}^0_1$ model for two benchmark points with 
$m_{\tilde\chi^0_1}=50$~GeV, as compared to official CMS-SUS-13-011 results given on \cite{cms-sus-13-011-twiki}. 
The next-to-last (last) line corresponds to the most sensitive
signal region for the benchmark point with $m_{\tilde t}=650$ (250) GeV as in the official CMS cut flow, while all other cuts are common to all signal regions targeting the $\tilde{t} \to t \tilde\chi^0_1$ decay mode. The uncertainties given for the CMS event numbers are statistical only.  In contrast to Tables~\ref{tab:NevStopbcharginoLowDeltaMMET2501} and \ref{tab:NevStopTneutralinoLowDeltaMMET2502}, no trigger efficiency or ISR reweighting is applied here. 
See \cite{cms-sus-13-011-twiki} for more details on the definition of the cuts.  
\label{tab:cms-13-011-cutflow}}
\begin{center}
\begin{tabular}{ l ||c|c||c|c}
\hline\noalign{\smallskip}
& \multicolumn{2}{|c||}{$m_{\tilde t}=650$~GeV} & \multicolumn{2}{c}{$m_{\tilde t}=250$~GeV}  \\
cut & CMS result & {\sc MA}\,5 result & CMS result & {\sc MA}5 result \\ 
\hline\noalign{\smallskip}
$1\ell\, + \ge 4{\rm jets} + E_T^{\rm miss}>50$~GeV & $31.6\pm0.3$ & $29.0$ & $8033.0\pm38.7$ &  $7365.0$  \\ 
+ $E_T^{\rm miss}>100$~GeV   & $29.7\pm0.3$ & $27.3$ & $4059.2\pm 27.5$  & $3787.2$ \\
+ $n_b\ge1$        & $25.2\pm0.2$ & $23.8$ & $3380.1\pm25.1$ & $3166.0$ \\
+ iso-track veto   & $21.0\pm0.2$ & $19.8$ & $2770.0\pm22.7$ & $2601.4$ \\
+ tau veto             & $20.6\pm0.2$ & $19.4$ & $2683.1\pm22.4$ & $2557.2$ \\
+ $\Delta\phi_{\rm min}>0.8$  & $17.8\pm0.2$ & $16.7$ & $2019.1\pm19.4$ & $2021.3$ \\
+ hadronic $\chi^2<5$  & $11.9\pm0.2$ & $9.8$ & $1375.9\pm16.0$ & $1092.0$ \\
+  $M_T>120$~GeV & $9.6\pm0.1$ & $7.9$ & $355.1\pm8.1$ & $261.3$ \\
${\rm high\,} \Delta M, E^{\rm miss}_T > 300~{\rm GeV}$ & $4.2\pm0.1$ & $3.9$ & --- & ---\\
${\rm low\,} \Delta M, E^{\rm miss}_T > 150~{\rm GeV}$ & --- & --- & $124.0\pm4.8$ & $107.9$\\
\noalign{\smallskip}\hline
\end{tabular}
\end{center}
\end{table*}

The validation of the reimplementation of the analysis can be done using the eleven benchmark points 
presented in the experimental paper: four for the ``T2tt'' simplified model (in which the stop always decays as $\tilde{t} \to t \tilde{\chi}^{0}_1$), and seven for the ``T2bW'' simplified model (in which the stop always decays as $\tilde{t} \to b \tilde{\chi}^{+}_1$), with different assumptions on the various masses. The distributions of the kinematic variables used in the analysis are given in Fig.~2 of~\cite{Chatrchyan:2013xna} after the preselection cuts, with at least one benchmark point for illustration. Also provided are the corresponding histograms after the \mbox{$M_T > 120$~GeV} cut, as supplementary material on the CMS Twiki page \cite{cms-sus-13-011-twiki}. We use this information, together with the final number of events in the individual SRs ({\it i.e.}, after all selection cuts) for given benchmark points provided in Tables~4 and 6 of~\cite{Chatrchyan:2013xna}. 

The validation material both before and after cuts defining the SRs is truly valuable information since one can separately check on the one hand the implementation of the kinematic variables and the preselection/cleaning cuts, 
and on  the other hand the series of cuts defining the SRs. Furthermore, the large number of benchmark points allows us to check in detail the quality of the reimplementation in complementary regions of phase space.

The validation process was based on (partonic) event samples, in LHE format~\cite{Boos:2001cv,Alwall:2006yp}, provided by the CMS collaboration. 
The provision of such event files greatly reduced the uncertainties in the first
stage of validation since it avoided possible differences in the configuration of the used
Monte Carlo tools. In the case of this CMS analysis, the setup of {\sc MadGraph}~5~\cite{Alwall:2011uj,Alwall:2014hca}---the event generator employed for generating the necessary hard scattering matrix elements---is crucial, in particular
with respect to the merging of samples with different (parton-level) jet multiplicities. 
The LHE files were passed through {\sc Pythia}~6.4~\cite{Sjostrand:2006za} for parton showering and hadronization, then processed by our modified version of \delphes~3 (see Section~\ref{sec:moddelphes3}) for the simulation of the detector effects. The number of events after cuts and histograms produced by \ma\ were then normalized to the correct luminosity after including cross sections at the next-to-leading
order and next-to-leading logarithmic (NLO+NLL) accuracy~\cite{Kramer:2012bx}, as tabulated by the LHC SUSY Cross Section Working Group~\cite{8tevxs_susy}.   

\begin{figure*}[!t]\centering
\includegraphics[width=6cm]{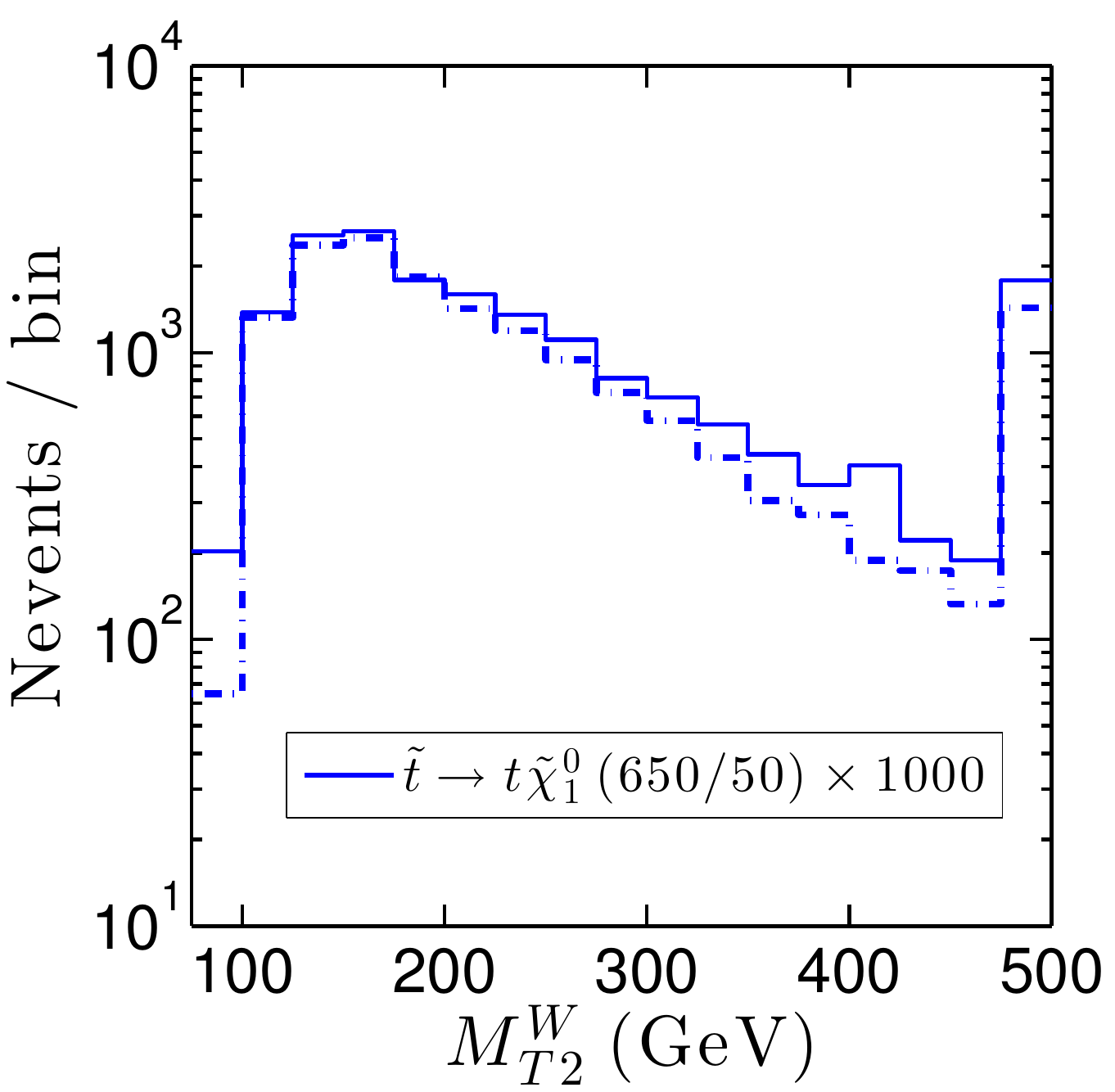}\quad
\includegraphics[width=6cm]{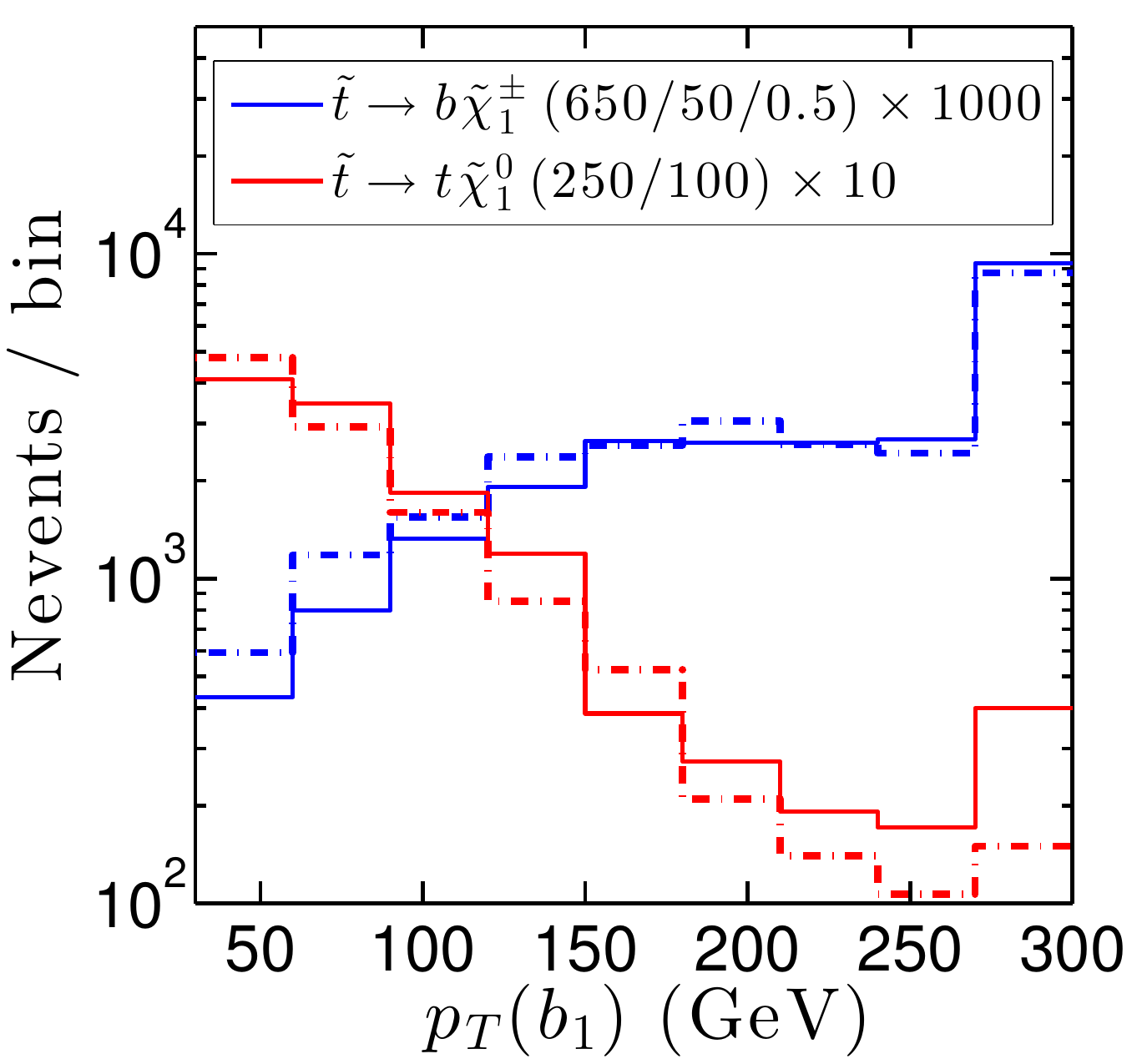}
\caption{Distributions of the kinematic variable $M^W_{T2}$ (left) and of the $p_T$ of the leading $b$-tagged jet (right) after the preselection cuts of the analysis CMS-SUS-13-011. The solid lines are obtained from our re-interpretation within \ma, while the dash-dotted lines correspond to the CMS results, given in Fig.~2 of~\cite{Chatrchyan:2013xna}. See captions of Tables~\ref{tab:NevStopbcharginoLowDeltaMMET2501} and \ref{tab:NevStopTneutralinoLowDeltaMMET2502} for the notation of the benchmark points.} \label{fig:kinvarsus13011}
\end{figure*}

Some examples of histograms reproduced for the validation are shown in Fig.~\ref{fig:kinvarsus13011}. The shapes of the distributions shown -- as well as all other distributions that we obtained but do not show here -- follow closely the ones from CMS, which indicates the correct implementation of the analysis and all the kinematic variables. 
(Note that discrepancies in bins where the number of events is relatively small, as seen on a logarithmic scale, suffers from larger statistical uncertainties and hence should not be over-interpreted.)
The expected yields for several benchmark points in their relevant SRs are given in Tables~\ref{tab:NevStopbcharginoLowDeltaMMET2501} and~\ref{tab:NevStopTneutralinoLowDeltaMMET2502}. 
The agreement is good for all tested benchmark points.

Upon our request, the CMS SUSY group furthermore provided detailed cut-flow tables, which are now also available at \cite{cms-sus-13-011-twiki}. 
These proved extremely useful because they allowed us to verify our implementation step-by-step in the analysis. 
A comparison of our results with the official CMS ones is given in Table~\ref{tab:cms-13-011-cutflow}. 
(Note that here no trigger efficiency or initial state radiation, ISR, reweighting is applied.)  
For both cases shown, CMS results are reproduced within about 20\%.  
On the whole, we conclude that our implementation gives reasonably accurate results 
(to the level that can be expected from fast simulation) and declare it as validated. 
As mentioned, the \ma\ code for this analysis, including extensive comments, is published as \cite{ma5code:cms-sus-13-011}.
More detailed validation material, including extra histograms and validation of the limit-setting procedure (see Section~\ref{sec:CLprocedure}), is available at~\cite{ma5wiki}.

%----------------------------------------------------------------------------------------------------------------------------------------------------------
\subsection{CMS-SUS-13-012: search for new physics through jet multiplicity and missing energy} 
\label{sec:cms-sus-13-012}
%----------------------------------------------------------------------------------------------------------------------------------------------------------

This CMS search for new physics in the hadronic activity in events with no leptons~\cite{Chatrchyan:2014lfa} targets a number of different signal topologies, in particular:
\begin{itemize}
\item gluino-pair production with $\tilde g\to q\bar q\tilde\chi^0_1$, denoted as T1qqqq topology in the following;
\item gluino-pair production with $\tilde g\to t\bar t\tilde\chi^0_1$, denoted as T1tttt; 
\item gluino-pair production with $\tilde g\to q\bar q\,\tilde\chi^0_2/\tilde\chi^\pm_1$, followed by 
$\tilde\chi^0_2,\chi^\pm_1 \to Z/W\tilde\chi^0_1$, generically denoted as T5VV; and 
\item squark-pair production with $\tilde q\to q\tilde\chi^0_1$, denoted as T2qq, 
\end{itemize}
following the CMS simplified models naming scheme \cite{Chatrchyan:2013sza}.

The analysis comprises 36 non-overlapping signal regions, each one defined as
a rectangular box volume in the space spanned by the variables $n_j$, $H_T$, and $\slashed{H}_T$. Here $n_j$ is the jet multiplicity of the event, $H_T$ is the scalar sum of the jet transverse momenta, and $\slashed{H}_T$ is the magnitude of the vector sum of the jets transverse momenta. Explicitly, 

\begin{equation}
  H_T = \sum_{\rm jets}{p_T}\,,\quad  \slashed{H}_T =  |\vec{\slashed{H}_T}| = \Big| \sum_{\rm jets}{{\bf p}_T} \Big|\,.
\label{eq:HTdefs}
\end{equation}

The event selection was primarily determined from the documentation in \cite{Chatrchyan:2014lfa}. This document describes six baseline selection criteria on the events, named MET Cleaning, No Lepton, $n_j>2$, $H_T>500$~GeV, $\slashed{H}_T>200$~GeV, and Min $\Delta\phi$({\rm jets}, $\vec{\slashed{H}_T}$). We note that the MET Cleaning cut involves a detailed consideration of spurious signals in the CMS detector, which we cannot simulate with \delphes. Instead, we simply multiply our event count by the efficiency given by CMS. (We stress again that such efficiencies being publicly available is extremely helpful.)

We validated the recast code against cut-flow tables and distributions of the kinematic variables provided by the CMS analysis team as per our request. The benchmark scenarios used are $(m_{\tilde g},\,m_{\tilde\chi^0_1})=(1100,\,125)$~GeV for the   T1qqqq, T1tttt and T5VV topologies, and $(m_{\tilde q},\,m_{\tilde\chi^0_1})=(700,\,100)$~GeV for the T2qq topology, 
with production cross sections of 10.2~fb and 63.4~fb, respectively~\cite{Kramer:2012bx,8tevxs_susy}.  
For the T5VV topology, one also needs the $\tilde\chi^\pm_1$ and $\tilde\chi^0_2$ masses; 
they are set to $612.5$~GeV for the $(m_{\tilde g},\,m_{\tilde\chi^0_1})=(1100,\,125)$~GeV benchmark point. 

The complete validation material from CMS is available in form of the PDF documents  
\emph{T1qqqq.pdf}, \emph{T1tttt.pdf}, \emph{T2qq.pdf} and \emph{T5VV.pdf} in the ``Attachments'' section 
on the analysis' wiki page~\cite{cms-sus-13-012-twiki}. 
These files correspond to the simplified SUSY models of the same names.  
For each of the four simplified-model scenarios, the CMS collaboration provided us with $10^5$ events in LHE format along with cut-flow tables and distributions in the variables $n_j$, $H_T$, and $\slashed{H}_T$ after each cut. As before, we passed these LHE files to {\sc Pythia}~6.4~\cite{Sjostrand:2006za} for showering and hadronization and finally to \delphes\ for detector simulation. 
The merging of the partonic events that exhibit different jet multiplicities was performed according to the setup read from the LHE files provided by CMS. 

A detail that required additional correspondence with the CMS analysis team were the pseudorapidity ($\eta$) cuts on the electrons and muons used for the lepton veto. We learned that the only requirement on these leptons is that %they have
$|\eta|<2.4$, and they are allowed to reside in the overlap region between the electromagnetic calorimeter barrel and the endcap.  
We also checked the dependence on the jet energy scale (JES) correction, which is set in the CMS \delphes\ card, 
to have good agreement in the $n_j$, $H_{T}$ and $\slashed{H}_T$ distributions, and found JES=1.0 to be optimal. 

\begin{figure}[t!]
\centering
\includegraphics[width=6.4cm]{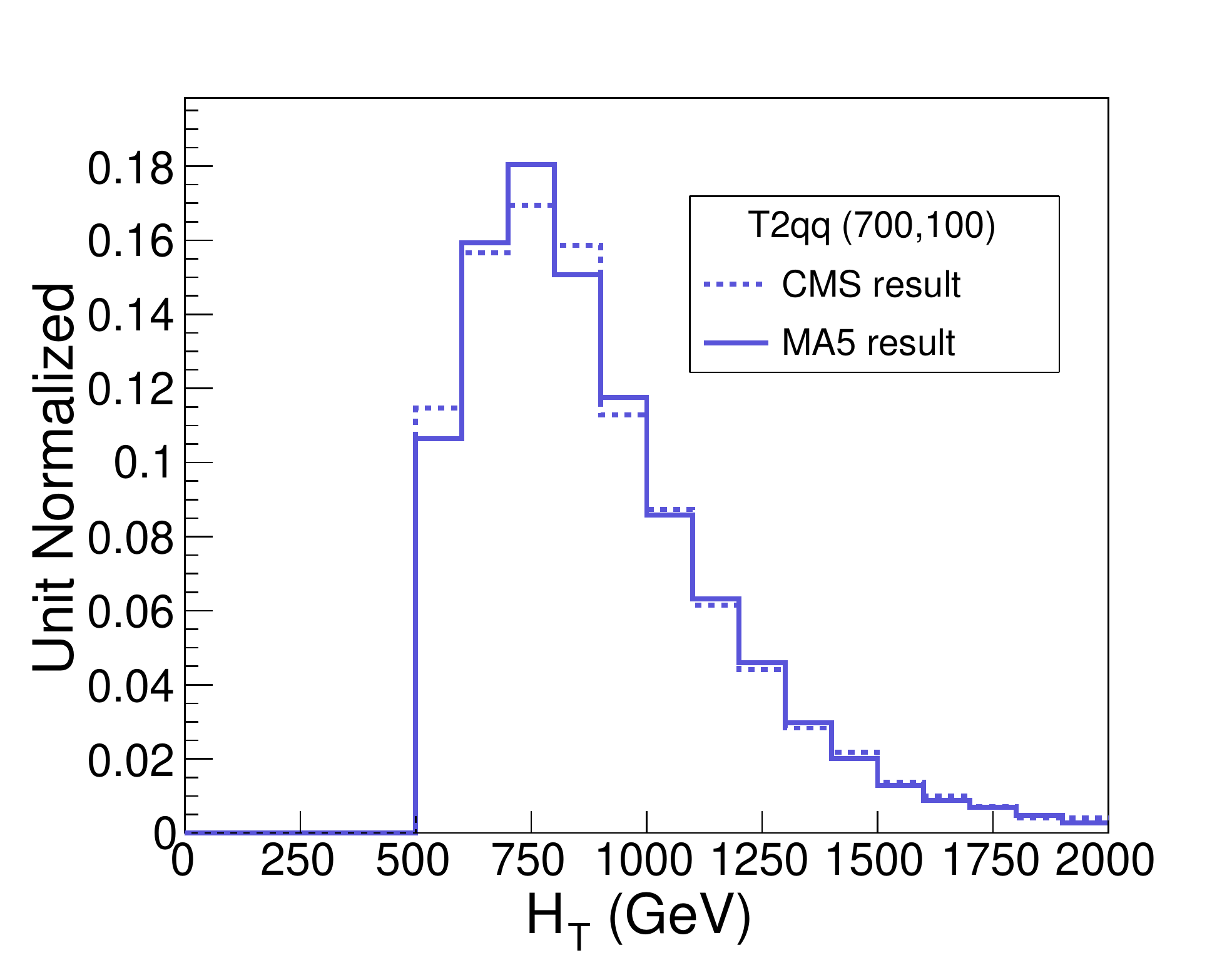}
\caption{Comparison between the official and \ma\ results  for the $H_T$ distribution after all baseline cuts, 
for the T2qq simplified model of CMS-SUS-13-012 with $(m_{\tilde q},\,m_{\tilde\chi^0_1})=(700,\,100)$~GeV.}
\label{fig:T2qqHT}
\end{figure}

\begin{figure}[t!]
\centering
\includegraphics[width=6.4cm]{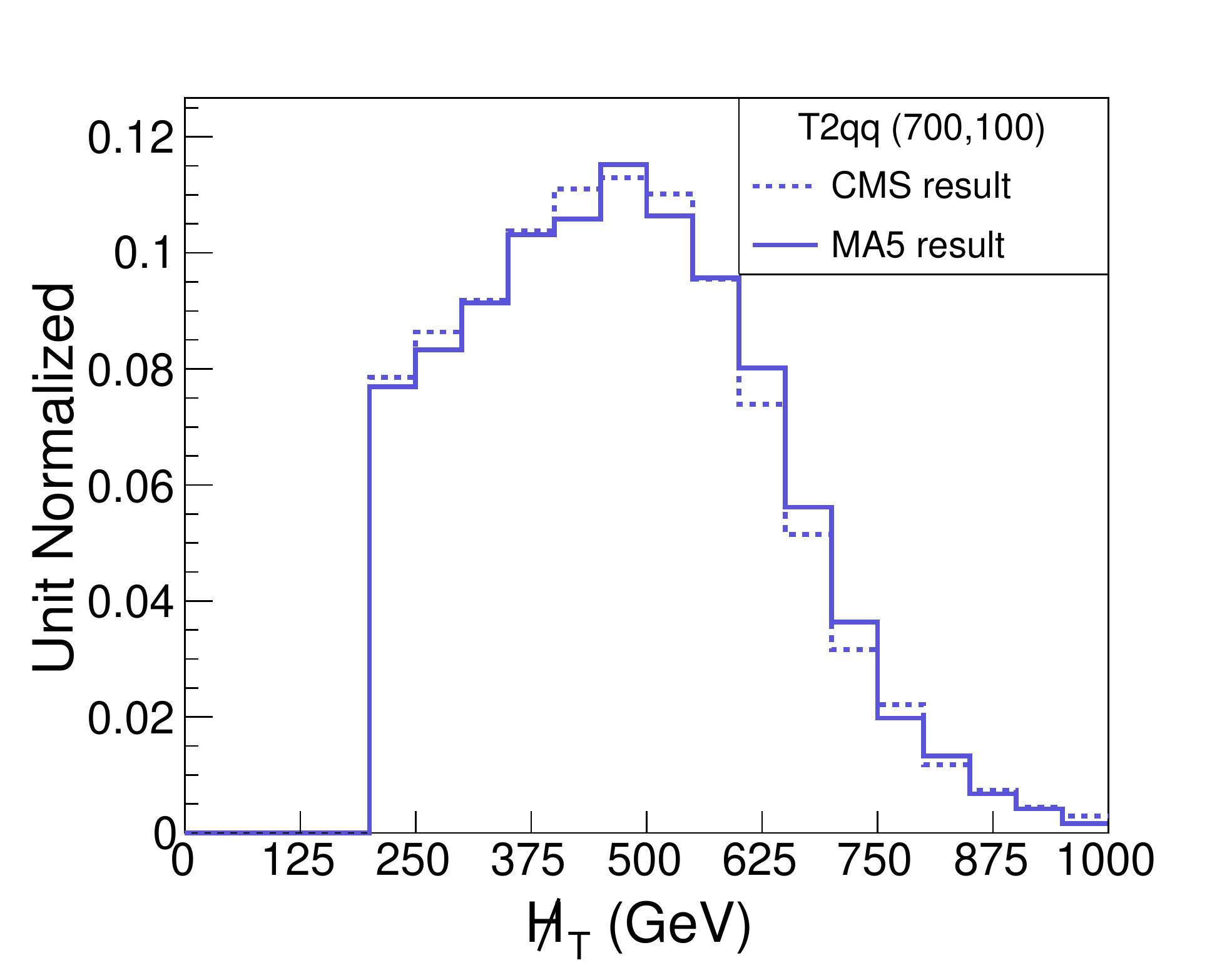}
\caption{Same as Fig.~\ref{fig:T2qqHT} but for the $\slashed{H}_T$ distribution.}
\label{fig:T2qqMHT}
\end{figure}

\begin{figure}[t!]
\centering
\includegraphics[width=6.4cm]{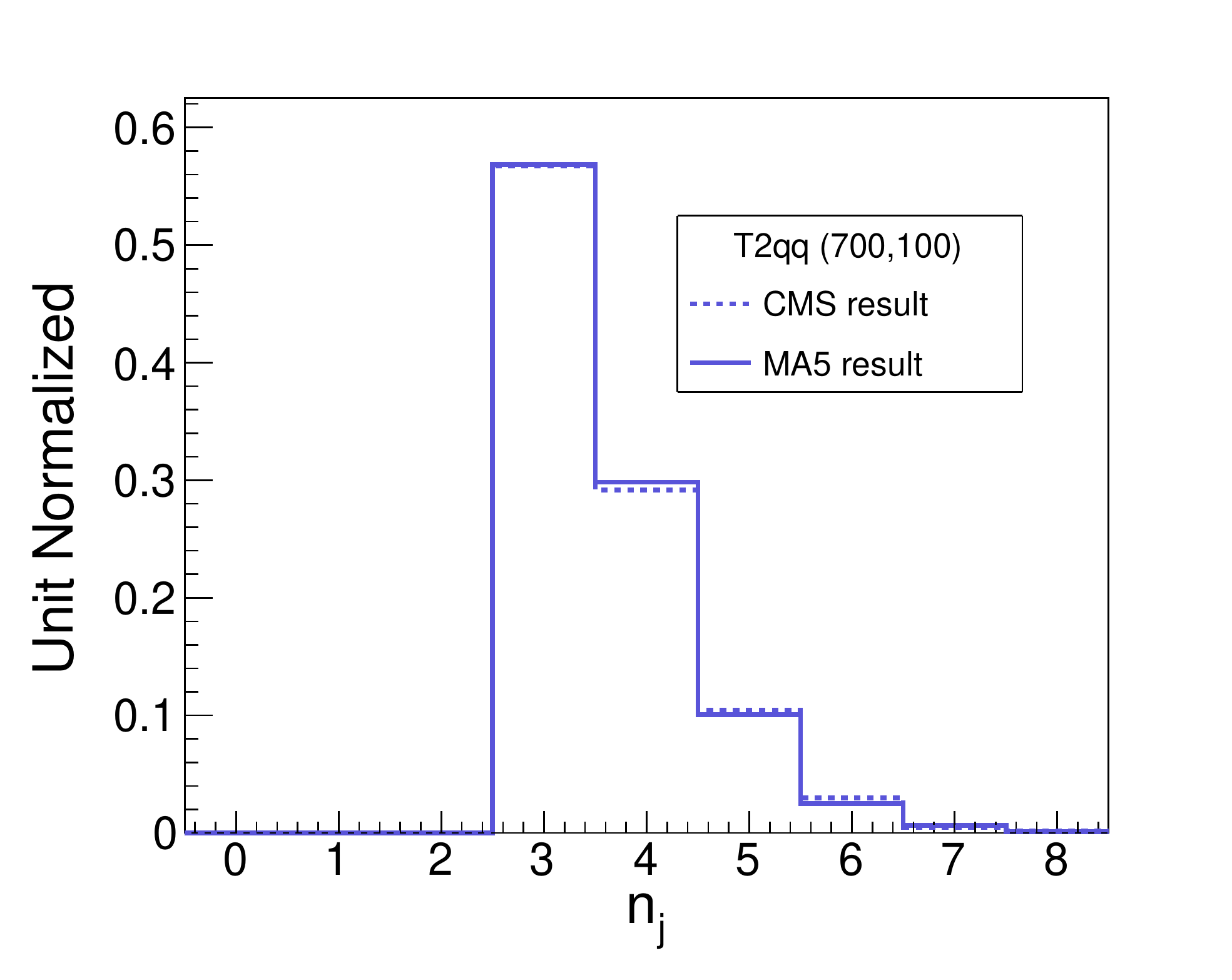}
\caption{Same as Fig.~\ref{fig:T2qqHT} but for the $n_j$ distribution.}
\label{fig:T2qqNJets}
\end{figure}

The results of our cut-flow counts for the various simplified models are shown alongside the official counts in 
Tables~\ref{table:CutFlowT1qqqq}  and \ref{table:CutFlowT5VV}. The results 
were obtained by normalizing with the cross section for each 
of the benchmark points and for an integrated luminosity of 19.3~fb$^{-1}$.
Moreover, some distributions after the baseline cuts for the case of the T2qq topology 
are shown in Figs.~\ref{fig:T2qqHT}--\ref{fig:T2qqNJets}. 
The distributions are normalized to unity and overlaid on the official plots obtained from the collaboration.

The agreement between the official and \ma\ results is better than 10\% throughout the baseline cut flows. The largest discrepancy arises from the lepton veto cut, which leads to a difference of up to about 5\% in the cut flow. The shapes of the distributions qualitatively match very well, and the peaking bins are in accordance with the official results. (This also holds for the other distributions not shown here for space considerations). 
The \ma\ implementation is available as \cite{ma5code:cms-sus-13-012}, and 
a detailed validation note comparing the recast results to the CMS ones can be found at \cite{ma5wiki}. 

\begin{table*}[!t]
\caption{Summary of yields for the baseline cuts for the T1qqqq and T1tttt topologies, as compared to the official CMS-SUS-13-012 results given on \cite{cms-sus-13-012-twiki}. The results are for the $(m_{\tilde g},\,m_{\tilde\chi^0_1})=(1100,\,125)$~GeV benchmark point.\label{table:CutFlowT1qqqq}}
\begin{center}
\begin{tabular}{ l ||c|c||c|c}
\hline\noalign{\smallskip}
& \multicolumn{2}{|c||}{T1qqqq} & \multicolumn{2}{c}{T1tttt}  \\
cut & CMS result & {\sc MA}\,5 result & CMS result & {\sc MA}5 result \\ 
\hline\noalign{\smallskip}
MET Cleaning & 190.6 & 190.6 & 190.5  & 190.5    \\ 
No Lepton   & 190.3 & 190.6  &  95.9 & 101.0 \\
+ $n_j>2$        & 188.1 & 188.5  & 95.8 & 100.9  \\
+ $H_T > 500$~GeV   & 187.6 & 188.1  & 95.1  &  100.0\\
+ $\slashed{H}_T >200$~GeV   &158.7  &  159.7 & 75.4 & 81.2  \\
+ Min $\Delta(\phi)$ & 130.8 & 131.1 &62.3 & 66.9 \\
\noalign{\smallskip}\hline
\end{tabular}
\end{center}
\end{table*}

\begin{table*}[!t]
\caption{Same as Table \ref{table:CutFlowT1qqqq}  but for the T2qq and T5VV topologies. 
The benchmark points used are $(m_{\tilde q},\,m_{\tilde\chi^0_1})=(700,\,100)$~GeV for T2qq and $(m_{\tilde g},\,m_{\tilde\chi^0_1})=(1100,\,125)$~GeV for T5VV.
\label{table:CutFlowT5VV}}
\begin{center}
\begin{tabular}{ l ||c|c||c|c}
\hline\noalign{\smallskip}
& \multicolumn{2}{|c||}{T2qq} & \multicolumn{2}{c}{T5VV}  \\
cut & CMS result & {\sc MA}\,5 result & CMS result & {\sc MA}5 result \\ 
\hline\noalign{\smallskip}
MET Cleaning & 1215.2 & 1215.2 &  189.9 & 189.9     \\ 
No Lepton   & 1212.8 & 1215.2  & 136.2  & 142.1  \\
+ $n_j>2$        & 675.9 & 691.5  &135.9  & 141.7  \\
+ $H_T > 500$~GeV   & 619.5 & 638.4   & 135.5  & 141.3 \\
+ $\slashed{H}_T >200$~GeV   & 524.0 & 539.6   & 108.8 & 115.2   \\
+ Min $\Delta(\phi)$ & 460.7 & 476.1  & 89.6 & 95.2 \\
\noalign{\smallskip}\hline
\end{tabular}
\end{center}
\end{table*}

%\clearpage

%----------------------------------------------------------------------------------------------------------------------------------------------------------
\subsection{CMS-SUS-13-016: search for gluinos in events with opposite-sign leptons, 
$b$-tagged jets and large missing energy} \label{sec:cms-sus-13-016}
%----------------------------------------------------------------------------------------------------------------------------------------------------------

The CMS analysis~\cite{CMS-PAS-SUS-13-016} searches for new physics in the multi-top final state. 
The primary target is gluino-pair production followed by $\tilde g\to t\bar t\tilde\chi^0_1$, \ie\ the 
T1tttt topology in the CMS simplified-model nomenclature. 
The dataset used corresponds to a total integrated luminosity of ${\cal L}=19.7$~fb$^{-1}$ at $\sqrt{s} = 8$ TeV. 

The analysis is not published yet but available as a Public Analysis Summary (PAS), 
which is overall well-documented. 
The signal selection requires two isolated leptons of opposite sign, a large number of jets, 
at least three $b$-tagged jets, and large missing transverse energy ($E^{\rm miss}_T>180$~GeV). 
Moreover, $|\eta|<1$ is required for the two leading jets. 
As there is only one SR, the exclusion is directly obtained from the upper limit on the number of events in the SR. 

Let us now turn to our \ma\ implementation and its validation. 
For the lepton isolation, we follow the same procedure as described above for CMS-SUS-13-011  
(see Section~\ref{sec:cms-sus-13-011}). 
Likewise, the $b$-tagging efficiency as function of $p_T$ is taken from~\cite{Chatrchyan:2013fea}. 
The most important piece of missing information in this PAS was  
a cut flow, which was however provided by the collaboration upon request and is now available 
on the analysis Twiki page~\cite{cms-sus-13-016-twiki}.

Along with the cut flows, CMS provided LHE files corresponding to two benchmark points 
for the T1tttt simplified model, one with 
$(m_{\tilde g},\,m_{\tilde\chi^0_1})=(1150,\,275)$~GeV, and one with 
$(m_{\tilde g},\,m_{\tilde\chi^0_1})=(1150,\,525)$~GeV.
The gluino-pair production cross section for these points is 6.7~fb with an uncertainty of 25\%~\cite{Kramer:2012bx,8tevxs_susy}.  
Unfortunately, these benchmark points differ by 25~GeV in the neutralino mass  from the ones used in the PAS, 
which have  $m_{\tilde\chi^0_1}=300$ and 500~GeV, respectively.
Although this is likely to induce some small differences in the event numbers and distributions, 
we chose to use the provided LHE files for validation because it avoids more important discrepancies due to 
differences in the configuration of the MC tools (\eg\ the exact version and setup of {\sc MadGraph} as well as 
the matching of parton-showers with hard scattering matrix elements and the merging of event samples 
exhibiting different jet multiplicities).\footnote{Note that having the exact same settings of the MC tools is important for  purposes of validation. A future user of the recast code, using \eg\ a different event  generator, may obtain a different result.}%will however be sensitive uncertainties  derived from this.} 
The LHE files were passed through {\sc Pythia}~6.4~\cite{Sjostrand:2006za} for parton showering and hadronization, 
with the correct merging parameters (given in the LHE files) taken into account. 
The detector simulation was then performed using the modified version of \delphes,  with the $b$-tagging efficiency taken from~\cite{Chatrchyan:2013fea} incorporated in the CMS card. 
The numbers of  events after all cuts were normalized using the cross section 
information tabulated by the LHC SUSY Cross Section Working Group 
and for an integrated luminosity of $19.7~\rm fb^{-1}$. 
Our cut flow is compared to the official CMS numbers in Table~\ref{tab:cms-13-016-cutflow}. 

\begin{figure}[!t]\centering
\vspace*{-2mm}
\includegraphics[width=6cm]{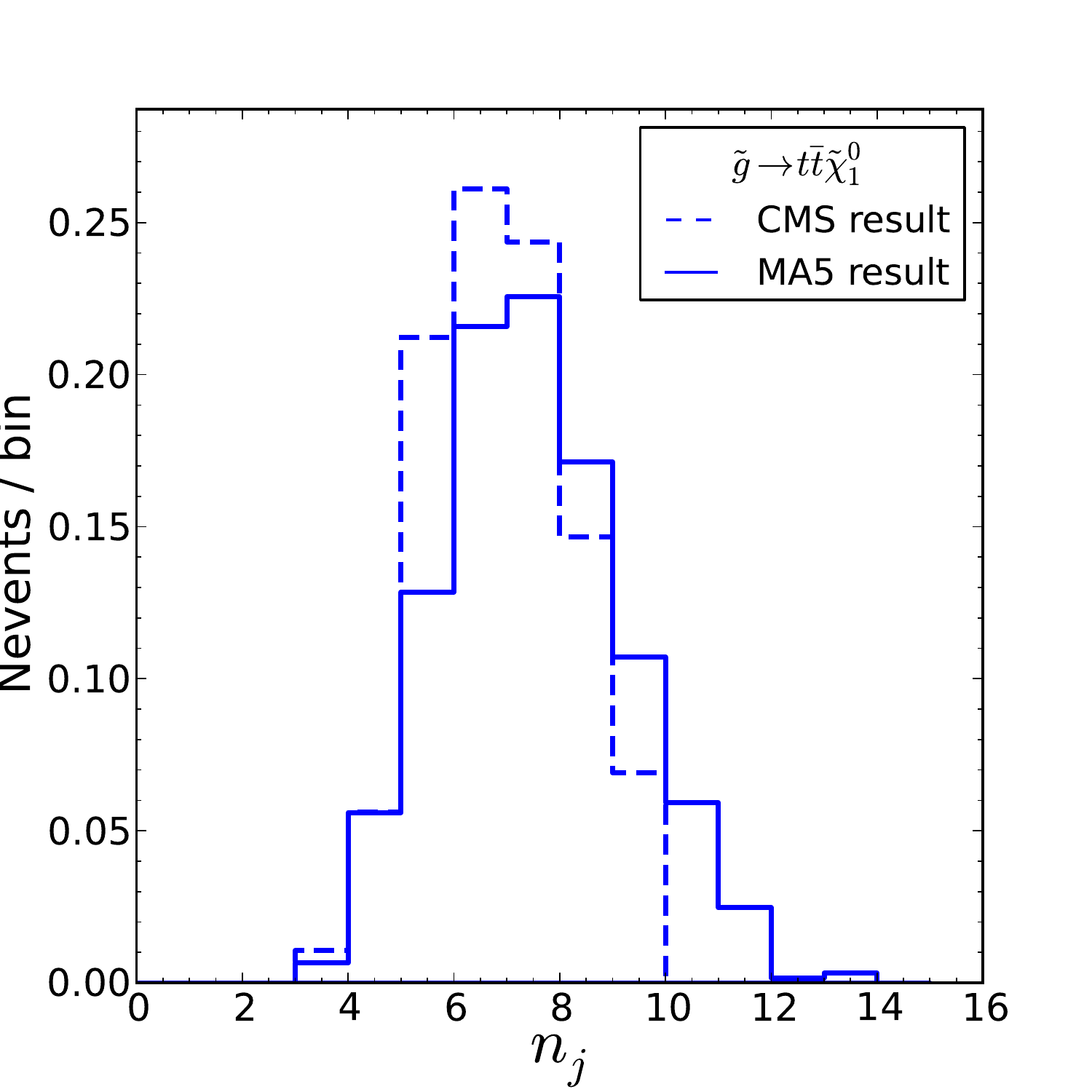} 
\caption{Distributions of the number of jets, $n_j$, corresponding to $(m_{\tilde g},\,m_{\tilde\chi^0_1})=(1150,\,275)$~GeV  for the analysis CMS-SUS-13-016. The dashed lines correspond to the CMS results, given in Fig.~1 of  Ref.~\cite{CMS-PAS-SUS-13-016}, while the solid lines are obtained from our  \ma\ implementation. Note that the plots are made by applying all cuts except the one represented.} \label{fig:sus13016njet}
\end{figure}

\begin{figure}[!t]\centering
\vspace*{-2mm}
\includegraphics[width=6cm]{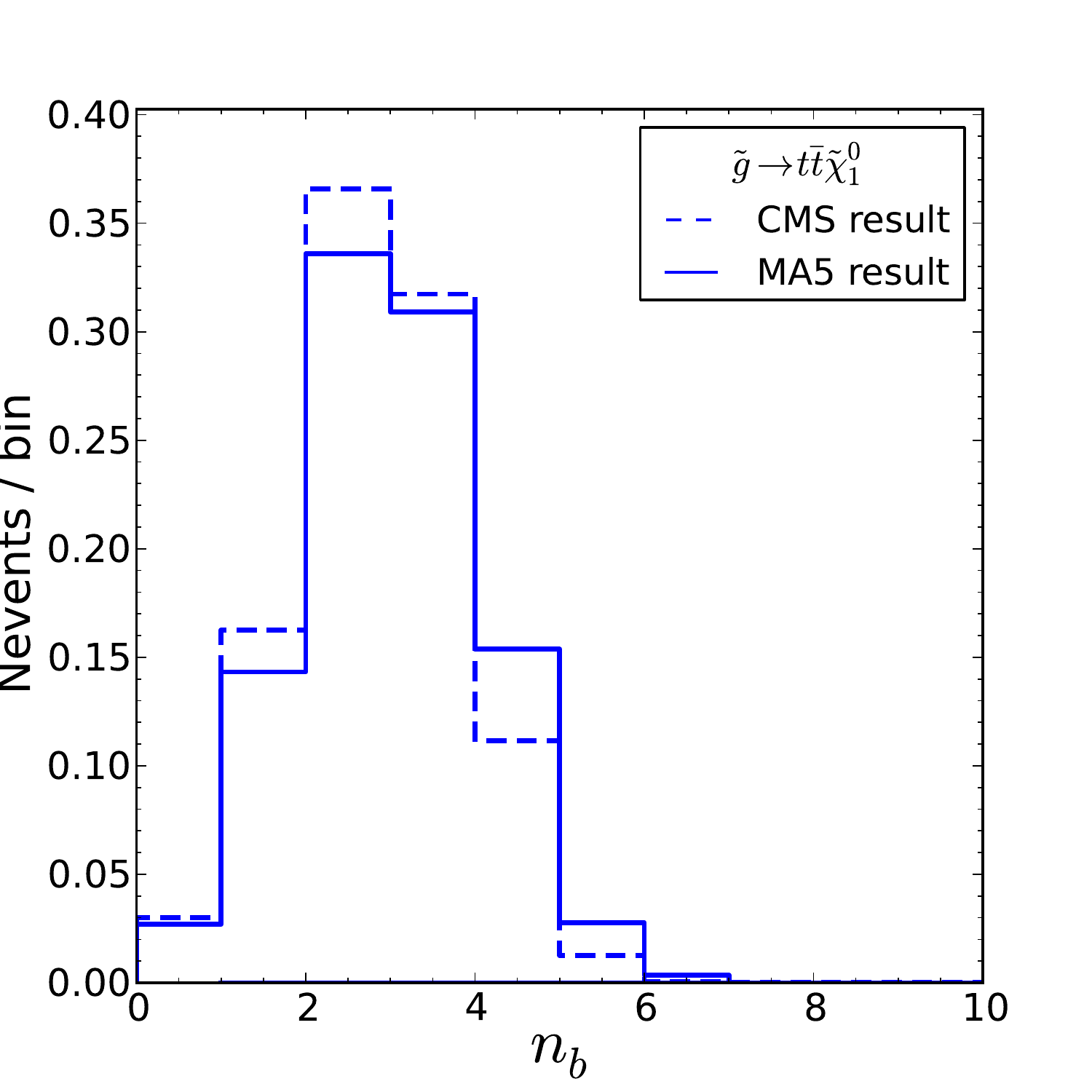}
\caption{As Fig.~\ref{fig:sus13016njet} but for the number of $b$-tagged jets, $n_b$.}
\label{fig:sus13016bjet}
\end{figure}

\begin{figure}[!t]\centering
\vspace*{-2mm}
\includegraphics[width=6cm]{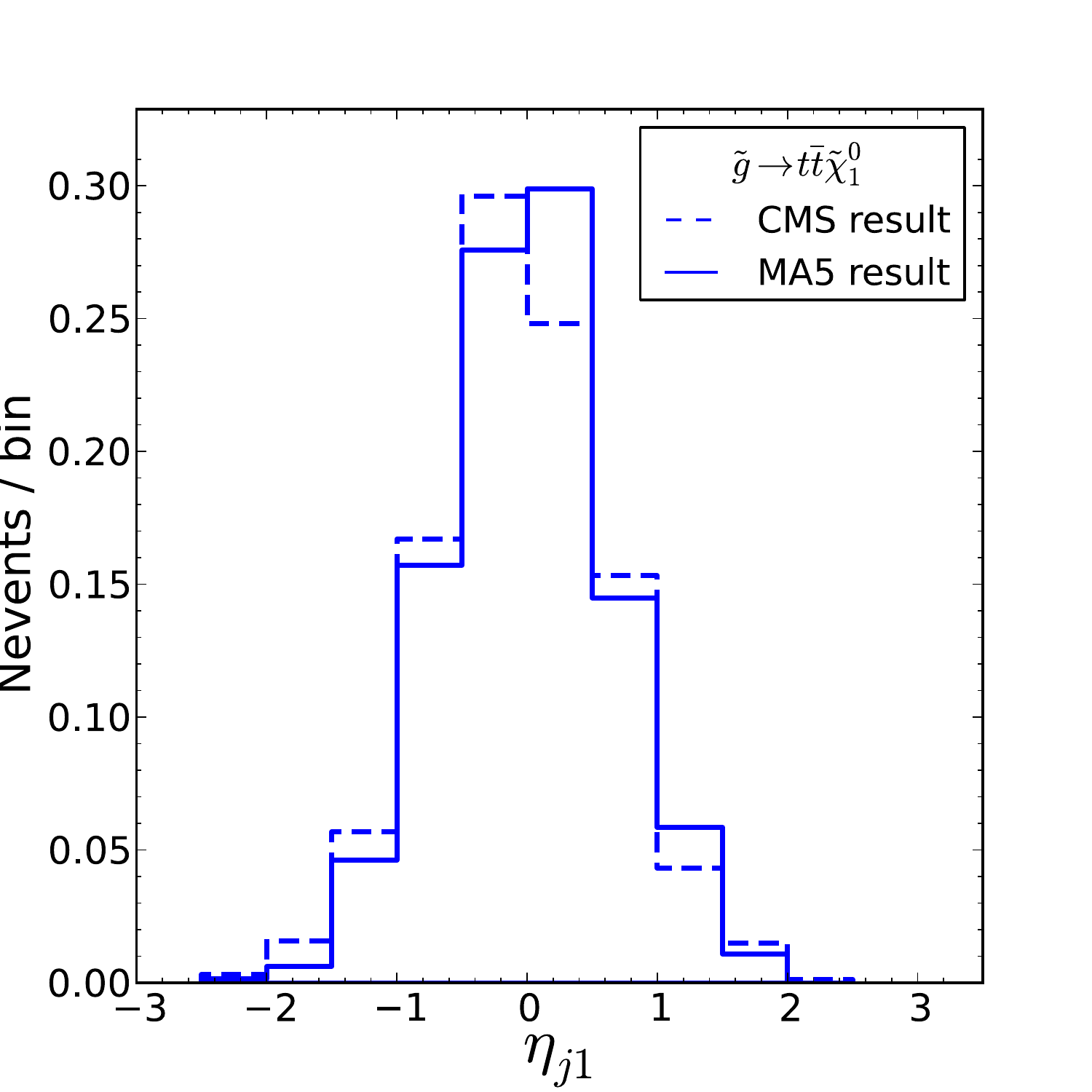}
\caption{As Fig.~\ref{fig:sus13016njet} but for the pseudorapidity of the leading jet, $\eta_{j1}$.}
\label{fig:sus13016eta1}
\end{figure}

\begin{figure}[!t]\centering
\vspace*{-2mm}
\includegraphics[width=6cm]{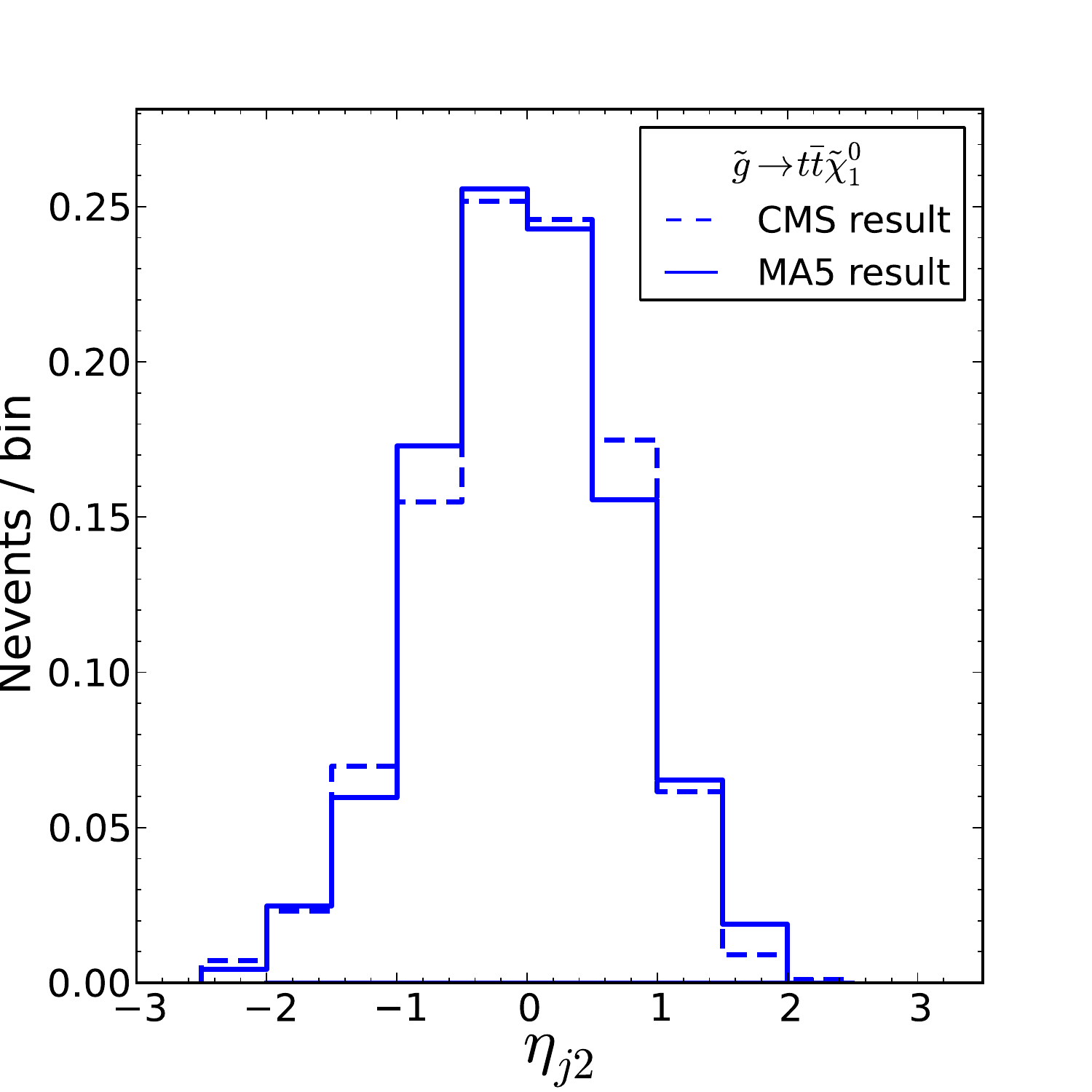}
\caption{As Fig.~\ref{fig:sus13016njet} but for the pseudorapidity of the sub-leading jet, $\eta_{j2}$.}
\label{fig:sus13016eta2}
\end{figure}

Figures~\ref{fig:sus13016njet}--\ref{fig:sus13016eta2} show  histograms of the kinematic selection variables for the 
$(m_{\tilde g},\,m_{\tilde\chi^0_1})=(1150,\,275)$~GeV benchmark point. 
Our \ma\ results are overlaid on the official results from Fig.~1 of \cite{CMS-PAS-SUS-13-016}, which we digitized. 
The plots were made by applying all cuts except the one represented, and 
all the histograms are normalized to unity. 
We note that the shapes of the distributions are in close agreement
with the official ones, with the exception of the $n_j$ 
distribution, which is slightly shifted towards higher jet multiplicity. 
Note also that the CMS histogram is cut off at $n_j = 10$, 
while the distribution in fact extends to higher  $n_j$. 

\begin{table*}[!t]
\caption{Summary of yields for the $\tilde{g} \rightarrow t \bar{t} \tilde{\chi}^0_1$ model for two benchmark points with 
$m_{\tilde{g}}=1150$~GeV, as compared to official CMS results given on \cite{cms-sus-13-016-twiki}. 
The uncertainties given for the CMS event numbers are statistical only. Note that the official numbers are available only
for $m_{\tilde\chi_{1}^{0}}=300$~GeV and 500~GeV. 
\label{tab:cms-13-016-cutflow}}
\begin{center}
\begin{tabular}{ l ||c|c||c|c}
\hline\noalign{\smallskip}
& \multicolumn{2}{|c||}{$m_{\tilde \chi_1^{0}}=275$~GeV} & \multicolumn{2}{c}{$m_{\tilde \chi_1^{0}}=525$~GeV}  \\
cut & CMS result & {\sc MA}\,5 result & CMS result & {\sc MA}5 result \\ 
\hline\noalign{\smallskip}
$2\ell\, + \ge 2{\rm jets} $ & $9.8\pm0.2$ & $9.0  $ & $ 9.5\pm 0.2$ &  $  8.9$  \\ 
+ $E_T^{\rm miss}>180$~GeV   & $ 7.5 $ & $7.3 $ & $  6.6$  & $ 6.4 $ \\
+ $n_j>4$        & $ 6.2 $ & $6.5 $ & $5.4  $ & $ 5.7 $ \\
+ $n_b>2$   & $2.6 $ & $3.1 $ & $ 2.3 $ & $ 2.6$ \\
+ $|\eta|_{j1}<1$  & $2.2 $ & $2.7 $ & $2.0 $ & $2.1 $ \\
+ $|\eta|_{j2}<1$  & $1.9 $ & $2.3 $ & $ 1.6 $ & $ 1.7$ \\
\noalign{\smallskip}\hline
\end{tabular}
\end{center}
\end{table*}

These differences can be attributed to various factors, one of which is the jet energy 
scale and resolution, for which a 8\% uncertainty is quoted in \cite{CMS-PAS-SUS-13-016}.  
Our results shown here were obtained with the JES parameter set to $1.0$ in the CMS \delphes\ card. 
A change of this parameter to $0.95$ does not change the results significantly, while a change to $0.9$ changes the final 
event count by 5 $\%$  after all cuts, and brings our $n_j$ distribution closer to the official one. 
Additionally there can be effects like pile-up or jet--lepton separation,  
which we cannot simulate reliably in this fast-simulation framework. 
Therefore we regard these effects as systematic uncertainties in our implementation.

Our final numbers of events for the two benchmark points agree within about 20\%  
with the official CMS numbers, see Table~\ref{tab:cms-13-016-cutflow}. This is well within 
the 17--39\% systematic uncertainty given in \cite{CMS-PAS-SUS-13-016}. 
Moreover, the individual cut efficiencies do not differ by more than 8\% for any cut 
for either of the benchmark points.  
This leads us to conclude that this implementation is well validated.  
The \ma\ code for this analysis is available as \cite{ma5code:cms-sus-13-016} 
and a detailed validation note is available on \cite{ma5wiki}.

%----------------------------------------------------------------------------------------------------------------------------------------------------------
\subsection{ATLAS-SUSY-2013-05: search for third-generation squarks 
in final states with zero leptons and two $b$-jets}
 \label{sec:atlas-susy-13-05}
%----------------------------------------------------------------------------------------------------------------------------------------------------------

In this ATLAS analysis \cite{Aad:2013ija}, stops and sbottoms 
are searched for in final states with 
large missing transverse momentum and two jets identified as $b$-jets. The
results are presented for an integrated luminosity of $20.1$~fb$^{-1}$ at
$\sqrt{s} = 8$ TeV. Two possible sets of SUSY mass spectra were
investigated in this analysis:
\begin{itemize}
 \item sbottom $\tilde{b}_1$ pair production with $\tilde{b}_1 \rightarrow b\tilde{\chi}_1^0$, and
 \item stop $\tilde{t}_1$ pair production with $\tilde{t}_1 \rightarrow b
\tilde{\chi}_1^\pm$, where the subsequent decay of the $\tilde{\chi}_1^\pm$ is
 invisible due to a small mass splitting with the $\tilde{\chi}_1^0$.
\end{itemize}
Two sets of SRs, denoted by SRA and SRB, are defined to provide sensitivity
to  the kinematic topologies associated with the two sets of mass
spectra. SRA  targets signal events with large mass splittings between the
squark and the neutralino by selecting two hard $b$-jets, while SRB is
designed to enhance the sensitivity when the squark--neutralino mass difference
is small by selecting a hard jet coming from ISR and two softer $b$-jets.

\begin{table*}[!t]
\caption{Summary of yields for SRA of ATLAS-SUSY-2013-05 corresponding to the benchmark points 
$(m_{\tilde b_1}, m_{\tilde \chi^0_1}) = (500,1)$~GeV and 
$(m_{\tilde t_1},m_{\tilde \chi^\pm_1}, m_{\tilde\chi^0_1})=(500,120,100)$~GeV,
as compared to official ATLAS results given on \cite{atlas-sus-13-05-twiki}. 
An $E^{\rm miss}_T$ filter is applied at the particle level. See \cite{atlas-sus-13-05-twiki} for more detail. 
\label{tab:atlas-13-05-cutflow-SRA}}
\begin{center}
\begin{tabular}{ l ||c|c||c|c}
\hline\noalign{\smallskip}
& \multicolumn{2}{|c||}{$m_{\tilde b_1}=500$~GeV} &
\multicolumn{2}{c}{$m_{\tilde t_1}=500$~GeV}  \\
cut & ATLAS result & {\sc MA}\,5 result & ATLAS result & {\sc MA}5 result \\ 
\hline\noalign{\smallskip} 
$E^{\rm miss}_T> 80$~GeV~filter & $1606.0$ & $1627.9$ & $1632.0$ & $1582.2$ \\ 
+ Lepton veto & $1505.0$ & $1592.6$ & $1061.0$ & $1140.8$ \\
+ $E^{\rm miss}_T > 150$~GeV & $1323.0$ & $1370.3$  & $859.0$ & $910.8$ \\ 
+ Jet Selection & $119.0$ & $122.2$ &$39.0$ & $39.6$ \\ 
+ $M_{bb} > 200$~GeV & $96.0$ & $99.3$ & $32.0$& $31.9$ \\ 
+ $M_{CT} > 150$~GeV & $82.0$ & $83.5$ & $26.8$ & $25.9$ \\
+ $M_{CT} > 200$~GeV & $67.0$ & $68.3$ & $20.2$ & $19.6$  \\
+ $M_{CT} > 250$~GeV & $51.0$ & $50.5$ & $13.2$ & $12.6$ \\
+ $M_{CT} > 300$~GeV & $35.0$ & $33.4$ & $7.7$ & $6.9$ \\
\noalign{\smallskip}\hline
\end{tabular}
\end{center}
\end{table*}

\begin{table*}[!t]
\caption{Summary of yields for SRB of ATLAS-SUSY-2013-05 corresponding to the benchmark points 
$(m_{\tilde b_1}, m_{\tilde \chi^0_1}) = (350,320)$~GeV and 
$(m_{\tilde t_1},m_{\tilde \chi^\pm_1}, m_{\tilde\chi^0_1})=(500,420,400)$~GeV,
as compared to official ATLAS results given on \cite{atlas-sus-13-05-twiki}. 
An $E^{\rm miss}_T$ filter is applied at the particle level. See \cite{atlas-sus-13-05-twiki} for more detail. 
\label{tab:atlas-13-05-cutflow-SRB}}
\begin{center}
\begin{tabular}{ l ||c|c||c|c}
\hline\noalign{\smallskip}
& \multicolumn{2}{|c||}{$m_{\tilde b_1}=350$~GeV} &
\multicolumn{2}{c}{$m_{\tilde t_1}=500$~GeV}  \\
cut & ATLAS result & {\sc MA}\,5 result & ATLAS result & {\sc MA}5 result \\ 
\hline\noalign{\smallskip} 
$E^{\rm miss}_T> 80$~GeV~filter  & $6221.0$  & $5990.6$ & $1329.0$ & $1109.9$\\ 
+ Lepton veto & $4069.0$ & $4773.4$  & $669.0$ & $816.5$  \\
+ $E^{\rm miss}_T > 250$~GeV & $798.3$  & $790.5$ & $93.0$ & $102.6$  \\
+ Jet Selection & $7.9$ & $7.2$ & $6.2$ & $4.7$ \\ 
+ $H_{T,3} < 50$~GeV & $5.2$  & $6.0$ & $3.0$ & $3.3$\\ 
\noalign{\smallskip}\hline
\end{tabular}
\end{center}
\end{table*}

For both SRs, events are selected by requiring a large amount of missing transverse energy,
\mbox{$E^{\rm miss}_T > 150$~GeV}, and any event containing an identified muon or electron is vetoed. 
For the SR selections, all jets with a pseudorapidity $|\eta| < 2.8$ are ordered according to their $p_T$, and 
two out of the $n$ selected jets are required to be $b$-tagged.

In the SRA, the first two leading jets must be $b$-tagged. The event is vetoed if any
additional central jet ($|\eta| < 2.8$) with $p_T > 50$~GeV is found. To reject
the multijet background, large $\Delta \phi_{\rm min}$ and $E^{\rm miss}_T/m_{\rm eff}$ 
are required.\footnote{$\Delta \phi_{\rm min}$ is the minimum azimuthal distance $\Delta \phi$ 
between any of the three leading jets and the $\mathbf{p}_T^{\rm miss}$ vector; 
$m_{\rm eff}$ is the scalar sum of the $p_T$ of the $k$
leading jets and the $E^{\rm miss}_T$, with $k=2$ for SRA and $k=3$ for SRB.}
To reduce the SM background, a cut on 
the invariant mass of the $b$-jet pair, $m_{bb}> 200$~GeV,  is applied. 
As a final selection, five different thresholds on the contransverse 
mass $m_{CT}$ \cite{Tovey:2008ui} ranging from 150~GeV to 350~GeV are demanded
to reduce backgrounds from top-quark production.\footnote{This peculiar kinematic variable 
is not yet implemented as a standard method in \ma. We thus used the public code computing
this quantity, including the correction due to ISR, available at 
\texttt{http://projects.hepforge.org/mctlib}.} 

In SRB, the sensitivity to small squark-neutralino mass difference is increased
by selecting events whose leading jet has a very large $p_T$, which is likely
to have been produced by ISR, recoiling against the squark-pair system. 
High thresholds on the leading jet and on the missing transverse
momentum, which are required to be almost back-to-back in $\phi$, are imposed.
The leading jet is required to be non-$b$-tagged and two additional jets are
required to be $b$-tagged. Just like for SRA, large values of $\Delta \phi_{\rm
min}$ and $E^{\rm miss}_T/m_{\rm eff}$ are required, thereby suppressing the
multijet background. The selection for SRB is finally completed by demanding
that the additional hadronic activity is bounded from above, $H_{T,3} < 50$~GeV. 
Here, $H_{T,3}$ is defined as the scalar sum of the $p_T$ of the jets, without including the three leading jets. 

\begin{figure*}[!t]
\centering
\includegraphics[width=5.2cm]{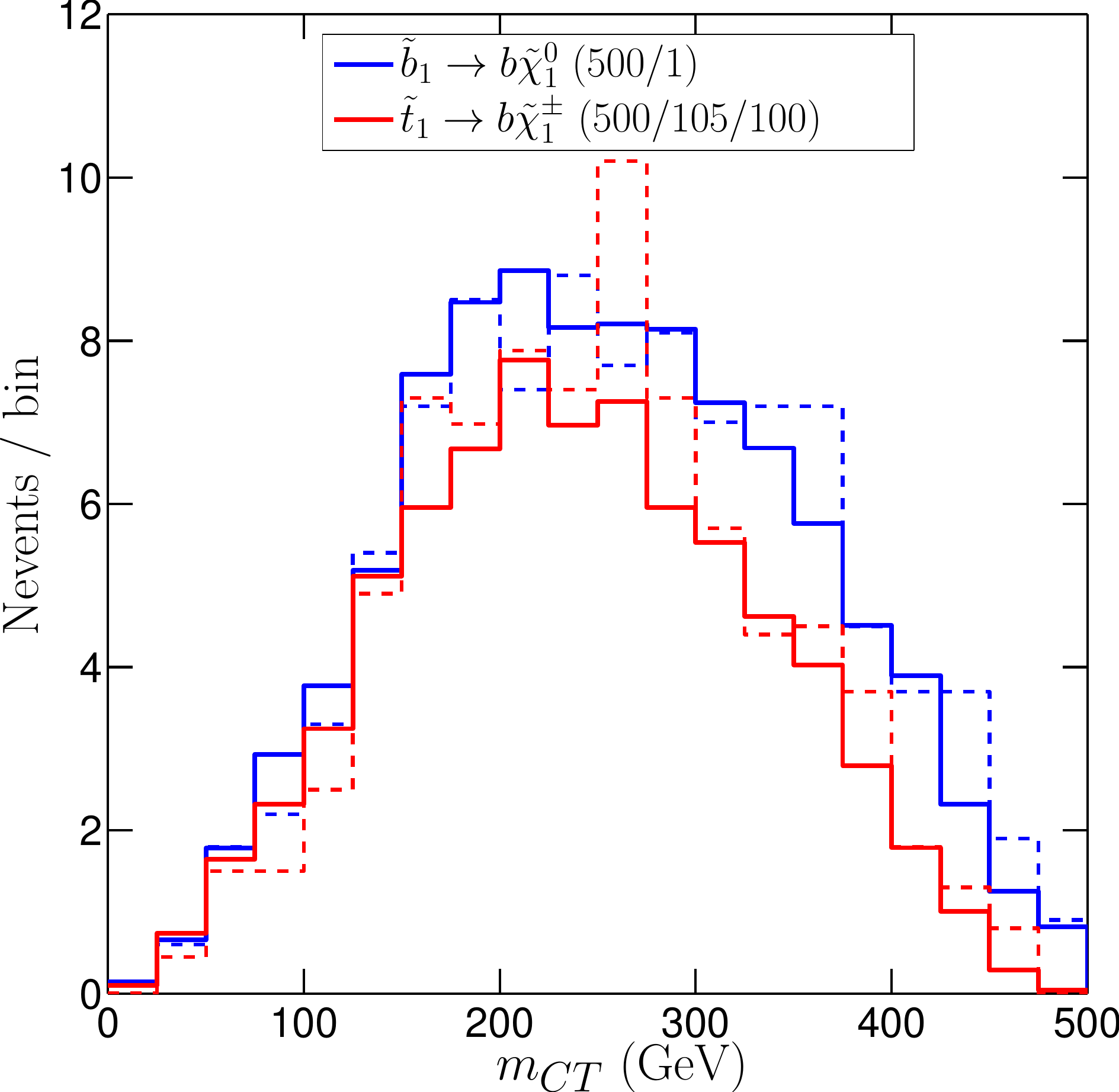} \qquad
\includegraphics[width=5.2cm]{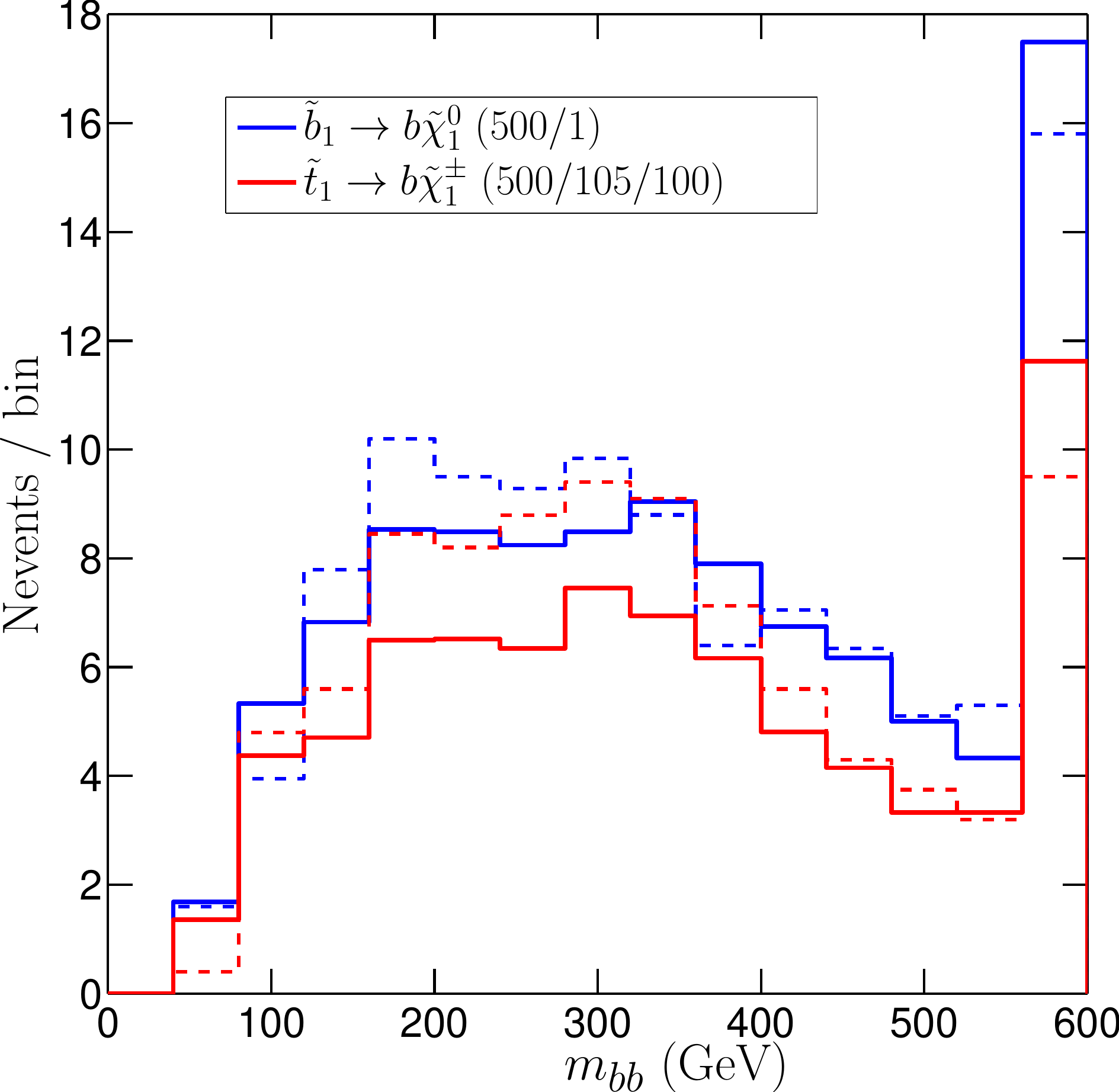}
\caption{\label{fig:SRAHistos}Distributions of $m_{CT}$ and of $m_{bb}$ for SRA of ATLAS-SUSY-2013-05 without their respective cut. The benchmark points used are $(m_{\tilde b_1},m_{\tilde\chi_1^0})$ = $(500,1)$~GeV (in blue) 
and $(m_{\tilde t_1},m_{\tilde \chi^\pm_1},m_{\tilde \chi^0_1})$ = $(500,105,100)$~GeV (in red). 
The solid lines correspond to our re-interpretation within \ma\ and the dashed lines to the ATLAS result.}
\end{figure*}

\begin{figure*}[!t]
\centering
\includegraphics[width=5.2cm]{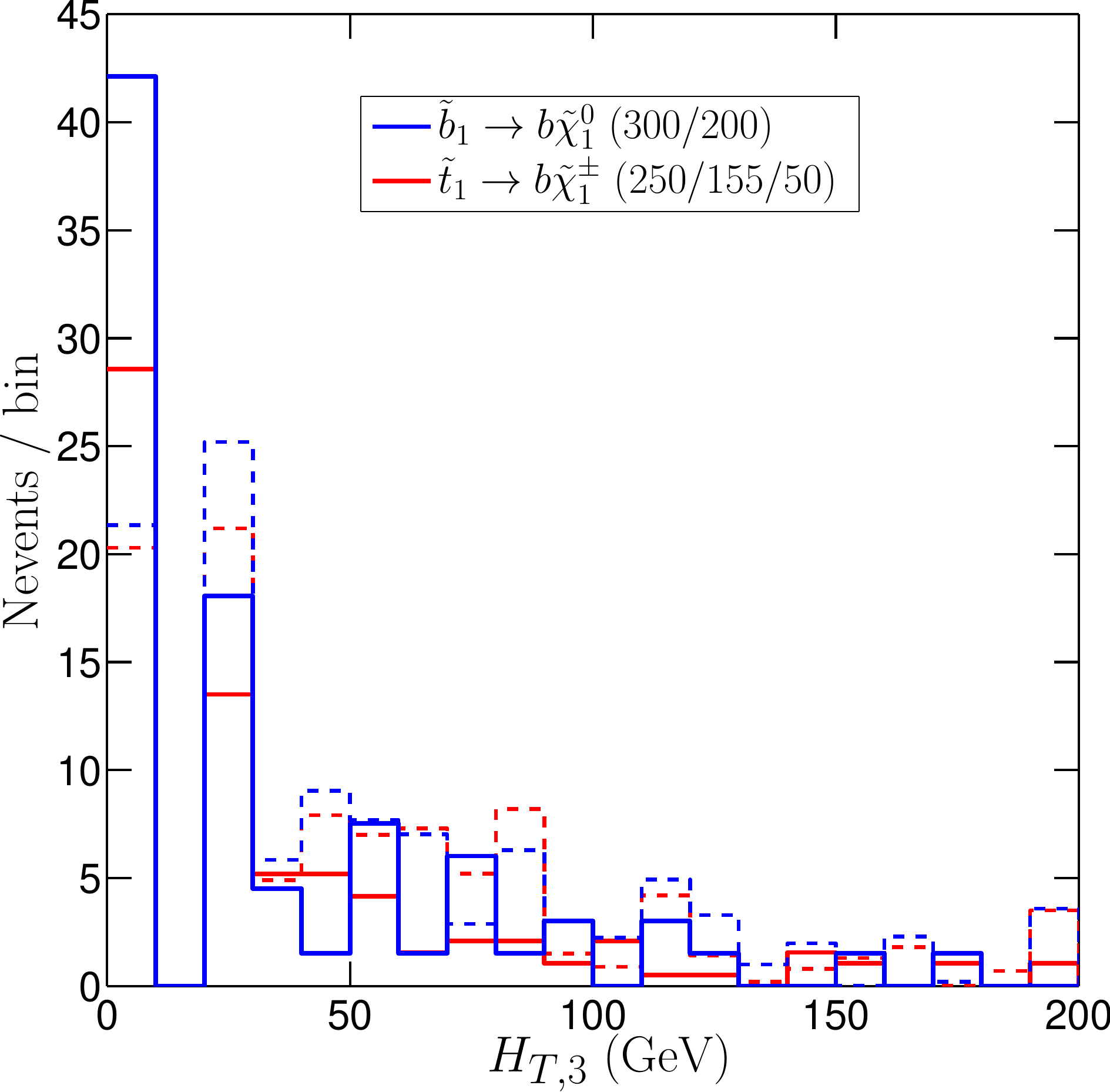} \qquad
\includegraphics[width=5.2cm]{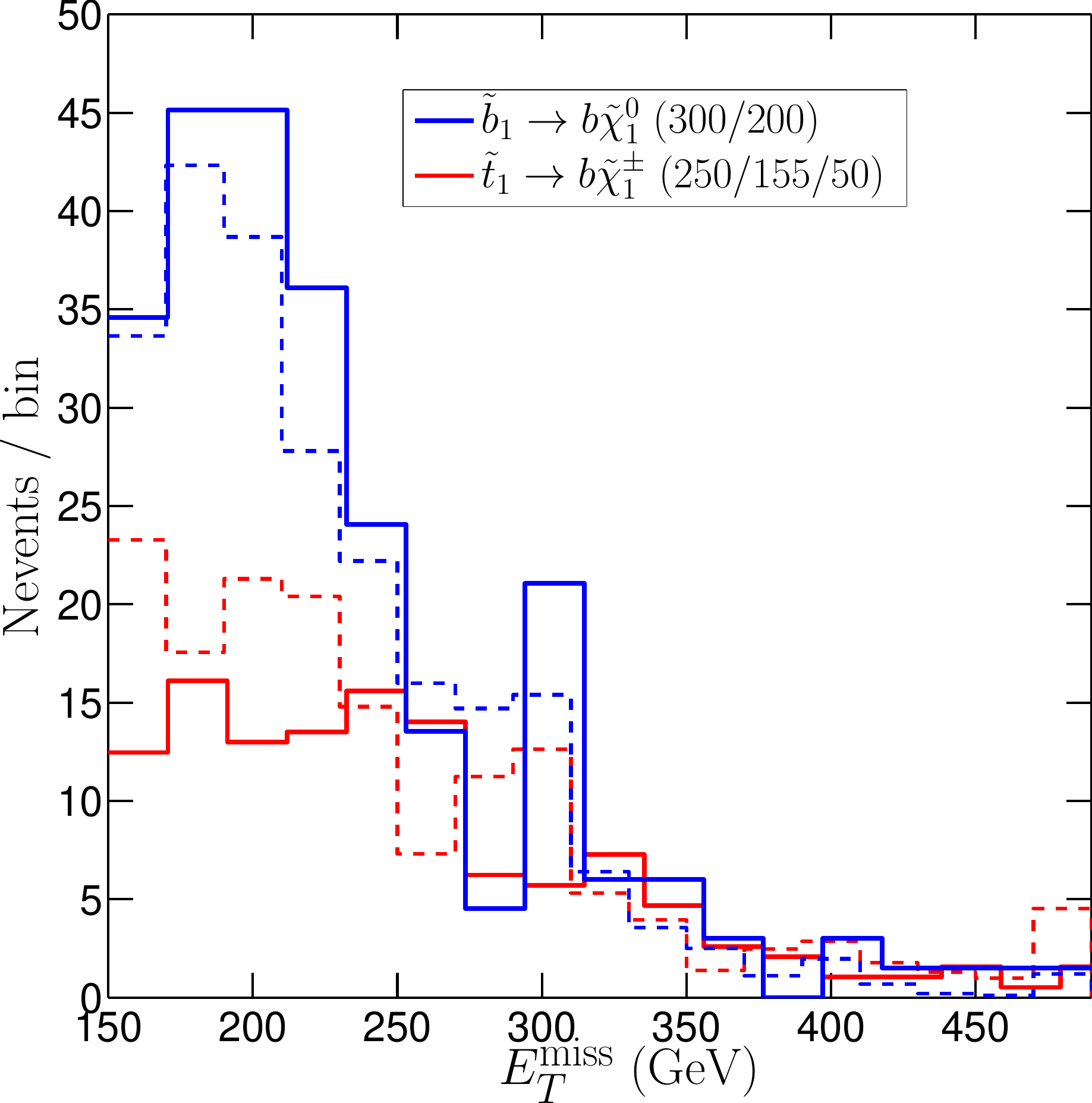}
\caption{\label{fig:SRBHistos}Distributions of $H_{T,3}$ and of $E_T^{\rm miss}$
for SRB of ATLAS-SUSY-2013-05 without their respective cut. The benchmark points
used are $(m_{\tilde b_1},m_{\tilde\chi_1^0})$ = $(300,200)$~GeV (in blue)
and $(m_{\tilde
t_1},m_{\tilde \chi^\pm_1},m_{\tilde \chi^0_1})$ = $(250,155,50)$~GeV (in red).
The solid lines correspond to our re-interpretation within \ma\ and the dashed
lines to the ATLAS result.}
\end{figure*}

The analysis is very well documented regarding physics, but for
recasting purposes more information than provided in the physics
paper~\cite{Aad:2013ija} and on the analysis Twiki page~\cite{atlas-sus-13-05-twiki}
was needed. 
Indeed this made the validation of the recast code seriously difficult 
in the earlier stages of the project. 
Since then, fortunately, two cut-flow tables were made public. 
Moreover, the ATLAS SUSY group provided us with general SUSY Les Houches Accord (SLHA)~\cite{Skands:2003cj} input files 
which we used to simulate the signal, as well as with the exact versions of the MC tools used to generate the SUSY samples,  which were not given in~\cite{Aad:2013ija}. 
When we simulated the signal samples with {\sc Madgraph\,5}~1.4.8~\cite{Alwall:2011uj,Alwall:2014hca} and {\sc
Pythia}~6.4~\cite{Sjostrand:2006za}, we introduced nonetheless additional sources 
of uncertainties since the complete MC configuration which was 
used for the signal simulation in ATLAS was not known. For example,  the
run card  for {\sc MadEvent}~\cite{Maltoni:2002qb} would be precious information.
Also, we are missing information on the trigger only and $b$-tagging efficiencies.

The comparison between the official cut flows and the ones obtained within \ma\ are presented 
in Tables~\ref{tab:atlas-13-05-cutflow-SRA} and \ref{tab:atlas-13-05-cutflow-SRB}.
The numbers  were normalized to $20.1$~fb$^{-1}$ of data using the cross sections
tabulated by the LHC SUSY Cross Section Working Group~\cite{Kramer:2012bx,8tevxs_susy}.
Overall the agreement is quite satisfactory, considering the expected accuracy
for a fast simulation. We observe the largest discrepancy in Table~\ref{tab:atlas-13-05-cutflow-SRB} 
in the final number of events in SRB after the
$H_{T,3}$ cut for the benchmark point $(m_{\tilde b_1},m_{\tilde \chi^0_1})$ =
$(350,320)$~GeV. This discrepancy will also be
exhibited in the histogram of the $H_{T,3}$ distribution.
In the analysis paper \cite{Aad:2013ija} there are four histograms of
distributions that we can compare against. For SRA, there are histograms of
$m_{CT}$ and of $m_{bb}$. Two benchmark
points are considered, $(m_{\tilde b_1},m_{\tilde\chi_1^0})$ = $(500,1)$~GeV and 
$(m_{\tilde t_1},m_{\tilde \chi^\pm_1},m_{\tilde \chi^0_1})$ = $(500,105,100)$~GeV, 
which are different from those used for the cut flows. 
There are also two such
distributions for SRB, the $H_{T,3}$ distribution and the missing
transverse energy $E_T^{\rm miss}$. The corresponding benchmark points are 
$(m_{\tilde b_1},m_{\tilde \chi_1^0})$ = $(300,200)$~GeV and 
$(m_{\tilde t_1},m_{\tilde \chi^\pm_1}, m_{\tilde \chi^0_1})$ = $(250,155,150)$~GeV.

As far as the SRA distributions are concerned, see Fig.~\ref{fig:SRAHistos}, 
the agreement between our recast analysis and the official one is very good. 

The situation is less satisfactory in the SRB case.
As already pointed out regarding the cut flow of
Table~\ref{tab:atlas-13-05-cutflow-SRB}, the treatment of the $H_{T,3}$
variable seems problematic; we indeed observe a large excess of events in the
very first bin of its distribution with respect to the official result from
ATLAS. The very first bin corresponds to events where there are no additional
jets ($H_{T,3}=0$~GeV) except the ones which are required to select the event.
The second bin is empty since jets are required to have $p_T > 20$~GeV. This
shows that, after detector simulation, we do no get enough jet activity. One
possible explanation for this might be that we do not account for pile-up effects. 
According to private communication with ATLAS, the
discrepancy is however too large to be accounted for by the pile-up only. 
Varying the JES by a fixed factor does not improve much the situation for the 
very first bin but can lead to improvement in the next bins of the $H_{T,3}$ distribution. 
However this also has an impact on the $E_T^{\rm miss}$ distribution, which 
gets significantly modified. A possible solution might be a parameterization of the 
JES in terms of the $p_T$ of the jets for these signal regions, since for low $p_T$ 
it may vary significantly. In any case, in \cite{Aad:2013ija}, the JES uncertainty was
carefully estimated and amounts to only 3\% in SRB. Last but not least, it
appears that, at the calorimeter level, \delphes\ undersmears jets (and thus
MET) compared to ATLAS.\footnote{We thank Jamie Tattersall for
pointing this fact out.} Therefore the $p_T$ distribution of soft 
jets is too sharp and the hadronic activity is reduced too much by the
$p_T > 20$~GeV cut. Moreover, for such jets with low $p_T$ the QCD uncertainties 
are substantial. To investigate the issue more deeply, a more detailed cut flow 
apportioning the ``Jet selection'' line in Table~\ref{tab:atlas-13-05-cutflow-SRB} would 
be helpful. 
%appreciable, since it directly impacts the $H_{T,3}$ variable.

We conclude that for SRA the agreement is quite good. For SRB the efficiency
of the $H_{T,3}$ cut differs from the official analysis by about 20\%, which is acceptable 
from a fast-simulation viewpoint. Moreover, according to \cite{Barnard:2014joa}
the sensitivity of SRB is difficult to reproduce while the analysis is generally
dominated by SRA, as can also be seen in Fig.~3 of the auxiliary figures of
\cite{atlas-sus-13-05-twiki}. Overall this leads us to
conclude that this implementation is validated to the best that could be done.
The recast code is available as \cite{ma5code:atlas-susy-2013-05}, and a 
detailed validation note can be found on \cite{ma5wiki}.

%----------------------------------------------------------------------------------------------------------------------------------------------------------
\subsection{ATLAS-SUSY-2013-11: search for charginos, neutralinos and leptons in di-lepton final states}
 \label{sec:atlas-susy-13-11}
%----------------------------------------------------------------------------------------------------------------------------------------------------------

We consider the ATLAS search for the electroweak production of 
charginos, neutralinos and sleptons in final sta\-tes with two leptons (electrons and muons) 
and missing transverse momentum based on $20.3$~fb$^{-1}$  of data at 8 TeV~\cite{Aad:2014vma}.
The event selection requires two signal leptons of opposite charge, with $p_T > 35$~GeV and $p_T > 20$~GeV. 
Two kind of final states are considered:  same flavor (SF = $e^+e^-$ or $\mu^+\mu^-$) and 
different flavors (DF = $e^\pm\mu^\mp$).

Three types of signal regions are defined in this analysis. First, the $m_{T2}$ and $WW$ signal regions
require the invariant mass of the lepton pair to be outside the $Z$ window, and jets are vetoed.
The $m_{T2}$ signal regions (SR~$m_{T2}$) target direct slepton-pair production and chargino-pair production
followed by slepton-mediated decays. 
Each $m_{T2}$ signal region is defined by its threshold on the $m_{T2}$ (``stransverse mass'') variable~\cite{Lester:1999tx,Cheng:2008hk} that is used for reducing the $t\bar t$ and $Wt$ backgrounds: $m_{T2} > 90$, $> 120$ and $> 150$~GeV, for SR-$m^{90}_{T2}$, SR-$m^{120}_{T2}$, and SR-$m^{150}_{T2}$, respectively. The implementation of this requirement is straightforward as the $m_{T2}$ variable is available as a standard method in \ma.

Next, the $WW$a, $WW$b and $WW$c signal regions (referred to as SR-$WW$) are designed to provide sensitivity
to $\tilde\chi_1^+\tilde\chi_1^-$ production followed by leptonic $W$ decays. Each of these three regions is optimized for a given kinematic configuration, using cuts on the invariant mass and/or transverse momentum of the slepton pair ($m_{\ell\ell}$ and $p_{T,\ell\ell}$, respectively), possibly combined with cuts on $m_{T2}$ and on the ``relative missing transverse momentum'' $E_T^{\rm miss,rel}$. Here, $E_T^{\rm miss,rel}$ is defined as the missing transverse momentum $E_T^{\rm miss}$ multiplied by $\sin \Delta \phi_{\ell,j}$ of the azimuthal angle between the direction of ${\bf p}_T^{\rm miss}$ and that of the closest lepton or jet, $\Delta \phi_{\ell,j}$, is below $\pi/2$. This modified $E_T^{\rm miss}$ aims at suppressing events where missing transverse momentum is likely to come from mis-measured jets and leptons.  

Finally, the $Z$jets signal region (SR-$Z$jets) targets $\tilde\chi^\pm_1 \tilde\chi^0_2$ production, followed by $\tilde\chi^\pm_1 \to W^\pm \tilde\chi^0_1$ and $\tilde\chi^0_2 \to Z \tilde\chi^0_1$, with 
hadronic $W$ and leptonic $Z$ decays. Unlike in the other regions, jets are not vetoed; instead at least two central ``light'' jets (non-$b$-tagged with $|\eta| < 2.4$) are required.
In addition to $m_{\ell\ell}$ being consistent with leptonic $Z$ decays, requirements are made on $E_T^{\rm miss,rel}$, $p_{T,\ell\ell}$, on the invariant mass of the two leading jets ($m_{jj}$) and on the separation between the two leptons ($\Delta R_{\ell\ell}$) in order to suppress, in particular, the $Z$ + jets background.

All signal regions separately consider SF and DF leptons, except SR-$Z$jets where only SF leptons are considered. In total, 20 potentially overlapping signal regions are defined (considering $ee$ and $\mu\mu$ signal regions separately, as required for comparison with the official ATLAS cut flows). Detailed electron efficiencies as a function of $p_T$ and $\eta$ are available in~\cite{ATLAS-CONF-2014-032}; we used the electron efficiencies as a function of $p_T$ for $|\eta| < 2.47$, while muon efficiencies were taken to be 100\% as a good approximation.
The analysis is very well-documented and gives clearly the various preselection criteria and signal region cuts. Moreover, an effort was made in the definition of the tested new physics scenarios: a whole section of the experimental publication is dedicated to the description of the different SUSY scenarios. Furthermore, SLHA files were uploaded to {\sc HepData} \cite{atlas-susy-11-hepdata} in May 2014 after discussion with the ATLAS SUSY conveners. 

For validation, at least one cut-flow table is given for every signal region and type of scenario tested, which is very good practice. In addition, several histograms are given and can be used to validate the distribution of, in particular, $E_T^{\rm miss,rel}$ and $m_{T2}$. Finally, regarding the interpretations in terms of simplified models, not only the information on the 95\%~confidence level (CL) upper bound on the visible cross section is given, but also the CL$_s$ value, which is useful for validation of the limit-setting procedure.
The only difficulty came from the benchmark points for direct slepton production. Given the SLHA files provided on {\sc HepData}, it was not clear whether the slepton masses given as $m_{\tilde\ell}$ in the cut-flow charts and histograms really correspond to the physical masses or to the slepton soft terms. The difference can be of several GeV, inducing some uncertainty in the kinematic distributions and in the production cross sections for these scenarios.

Event samples used for the validation were generated with {\sc Herwig++}~2.5.2~\cite{Bahr:2008pv}, using as input the SLHA files provided on {\tt HepData}. For each of the nine benchmark points we considered, $10^5$ events were generated. In the case of chargino-pair production, non-leptonic decays of the intermediate $W$-boson were filtered to increase statistics. Similarly, for chargino--neutralino production, non-leptonic decays of the intermediate $Z$-boson were filtered. The cross sections for the benchmark points, evaluated at the NLO+NLL accuracy~\cite{Fuks:2012qx,Fuks:2013vua,Fuks:2013lya},  were taken from the {\sc HepData} entry.

\begin{figure*}[!th]\centering
\includegraphics[width=5.9cm]{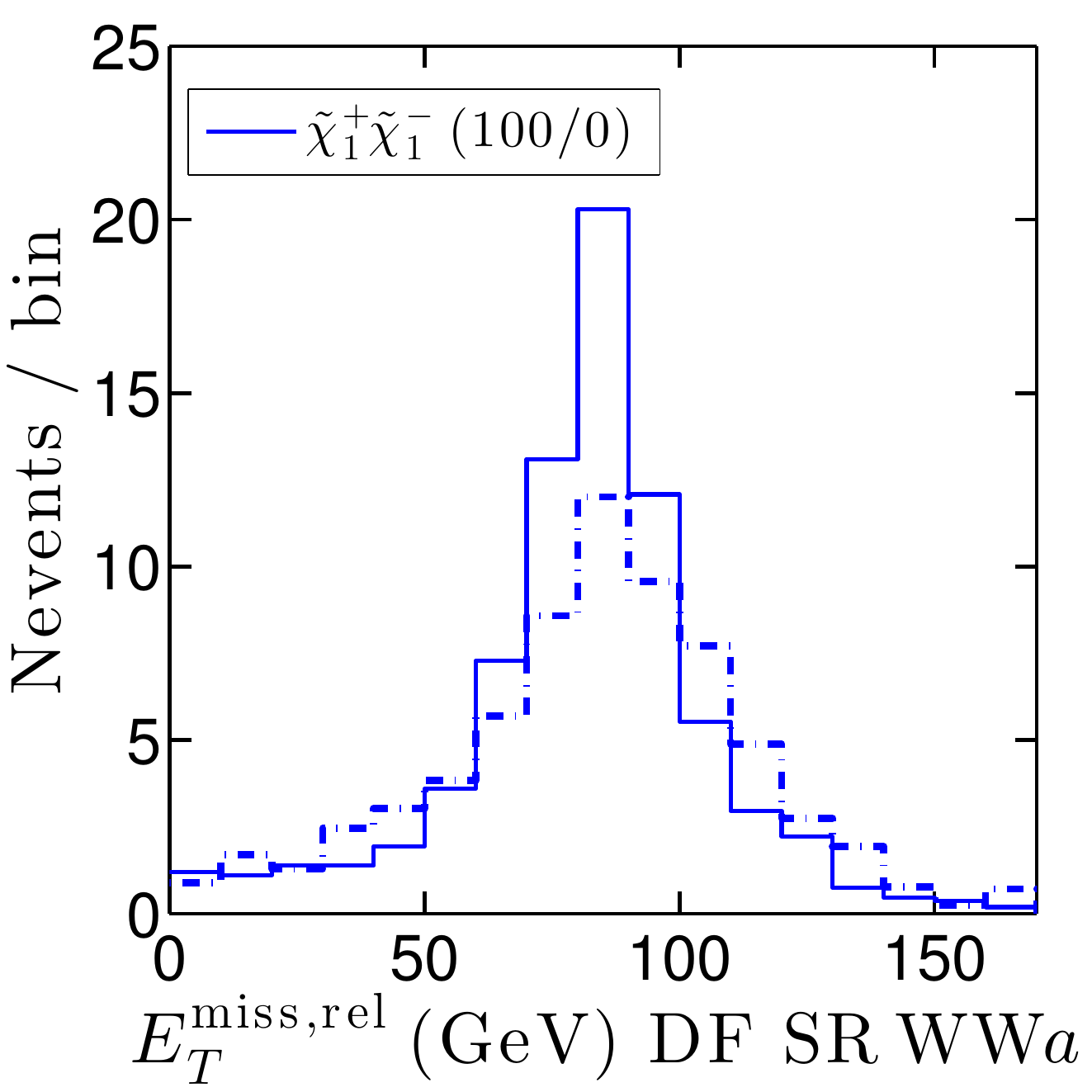}\qquad
\includegraphics[width=6cm]{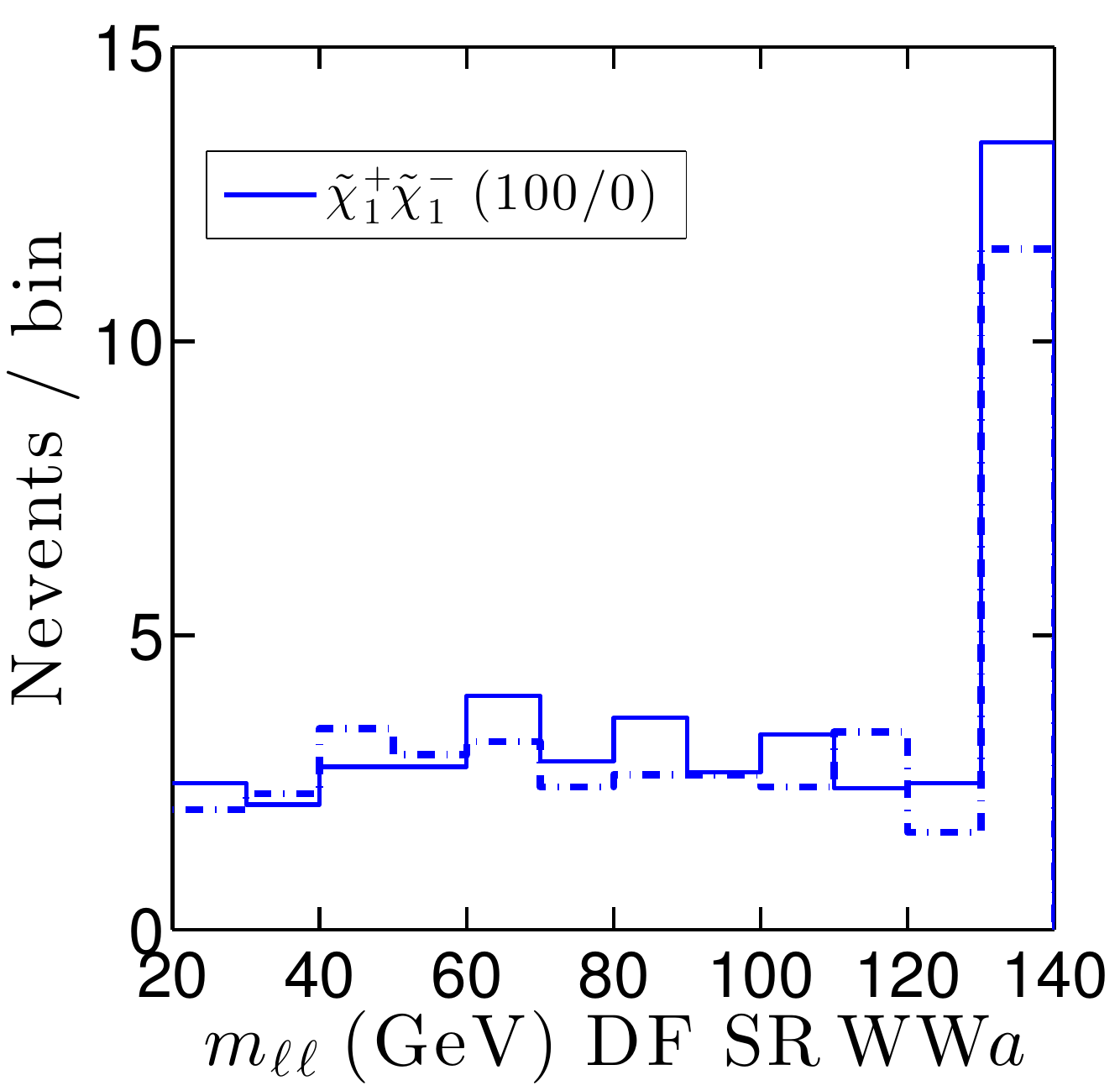}
\vspace*{1mm}
\caption{Distributions of $E_T^{\rm miss,rel}$ (left) and $m_{\ell\ell}$ (right) in the DF SR~$WW{\rm a}$ 
of ATLAS-SUSY-2013-11, for the benchmark point with $(m_{\tilde{\chi}^{\pm}_1}, m_{\tilde{\chi}^0_1})=(100,0)$~GeV,  
after all cuts except the ones on $m_{\ell\ell}$ and on $E_T^{\rm miss,rel}$ (left), or all cuts except the one on $m_{\ell\ell}$ (right). The solid lines are obtained from our re-interpretation within \ma, while the dash-dotted lines correspond to the official ATLAS results in~\cite{Aad:2014vma}.} 
\label{fig:atlas-11-figA}
\end{figure*}

\begin{figure*}[!th]\centering
\includegraphics[width=6cm]{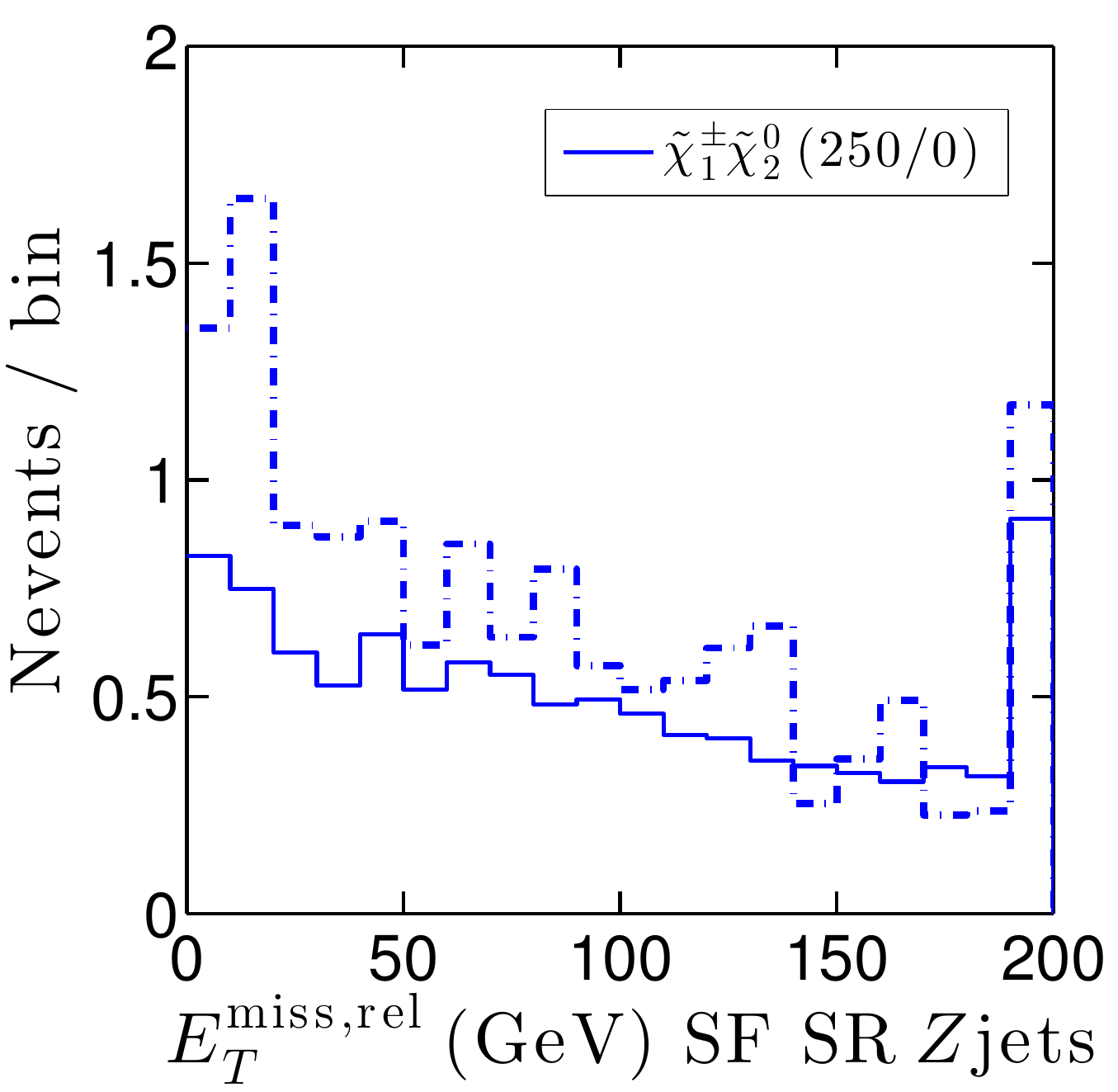}\qquad
\includegraphics[width=6cm]{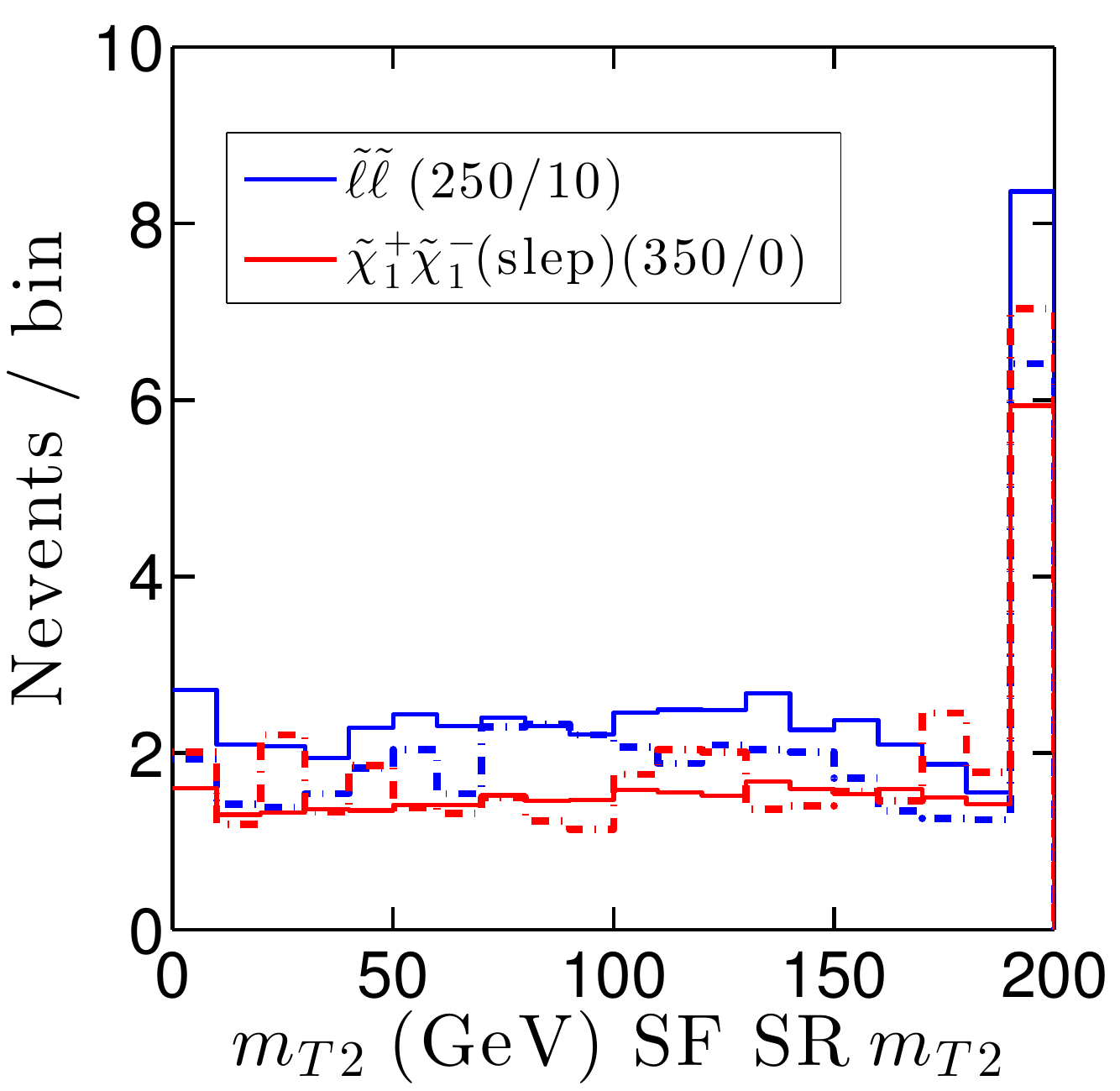}
\vspace*{1mm}
\caption{Distributions of $E_T^{\rm miss,rel}$ in the SF SR~$Z{\rm jets}$ (left) and $m_{T2}$ in the SF SR~$m_{T2}$ (right)  of ATLAS-SUSY-2013-11, after all cuts except the one on the variable plotted. The solid lines are obtained from our re-interpretation within \ma, while the dash-dotted lines correspond to the official ATLAS results in~\cite{Aad:2014vma}.} 
\label{fig:atlas-11-figB}
\end{figure*}

Tables~\ref{tab:cutflowWWaee_C1C1noslep1000}--\ref{tab:cutflowMT2120ee_slep25010} 
give some examples of cut flows for different benchmark points and signal regions, comparing 
the results obtained with our \ma\ implementation to the official ATLAS numbers. 
(The complete list of cut flows for all nine benchmark points is available at~\cite{ma5wiki}.)
We systematically find the jet veto to be less efficient than it should be, but did not find
any explanation for this effect. This was also noted in~\cite{Drees:2013wra}.
Still, reasonably good agreement is observed for the available benchmark points.
Distributions of $E_T^{\rm miss,rel}$, $m_{\ell\ell}$ and $m_{T2}$ in some signal regions are shown in Figs.~\ref{fig:atlas-11-figA} and \ref{fig:atlas-11-figB}. Good agreement is observed. Note that the fluctuations in the ATLAS results in the left panel of Fig.~\ref{fig:atlas-11-figB} may correspond to statistical fluctuations and/or uncertainties when digitizing the ATLAS histogram (the results are extracted from a logarithmic scale that spans over six orders of magnitude).

We conclude that our \ma\ implementation of ATLAS-SUSY-2013-11 reproduces well the experimental results. 
Our C++ code for this analysis is published as~\cite{ma5code:atlas-susy-2013-11}; complete validation materials including validation of the limit-setting procedure (see next section) can be found at~\cite{ma5wiki}.

\begin{table}[!t]
\caption{Cut flow for chargino-pair production in SR-$WW{\rm a}\,ee$ of ATLAS-SUSY-2013-11, 
for the benchmark point with $(m_{\tilde{\chi}^{\pm}_1}, m_{\tilde{\chi}^0_1})=(100,0)$~GeV.}
\label{tab:cutflowWWaee_C1C1noslep1000}
\begin{center}
\begin{tabular}{l|c|c}
cut & ATLAS result & {\sc MA}5 result \\ 
\hline
Initial number of events & & $12301.5$  \\ 
2 OS leptons & & $1666.5$  \\ 
$m_{\ell\ell} > 20$~GeV & & $1637.5$  \\ 
$\tau$ veto & & $1637.5$  \\ 
$ee$ leptons & $402.1$ & $392.9$ \\ 
jet veto & $198.6$ & $257.0$  \\ 
$Z$ veto & $165.0$ & $215.9$ \\ 
$p_{T,\ell\ell} > 80$~GeV & $28.0$ & $35.3$  \\ 
$E_T^{\rm miss,rel} > 80$~GeV & $14.7$ & $18.9$  \\ 
$m_{\ell\ell} < 120$~GeV & $9.2$ & $10.1$  \\ 
\hline 
\end{tabular}
\end{center}
\vspace*{-5mm}
\end{table}

\begin{table}[!t]
\caption{Cut flow for $\tilde{\chi}^{\pm}_1\tilde{\chi}^0_2$ associated production 
in SR-$Z{\rm jets}\,\mu\mu$ of ATLAS-SUSY-2013-11, 
for the benchmark point with $(m_{\tilde{\chi}^{\pm}_1}, m_{\tilde{\chi}^0_1})=(350,50)$~GeV.}
\label{tab:cutflowZjetsmumu_C1N235050}
\begin{center}
\begin{tabular}{l|c|c}
cut & ATLAS result & {\sc MA}5 result \\ 
\hline
Initial number of events & & $152.2$  \\ 
2 OS leptons & & $47.0$ \\ 
$m_{\ell\ell} > 20$~GeV & & $46.9$  \\ 
$\tau$ veto & & $46.9$  \\ 
$\mu\mu$ leptons & $16.4$ & $24.2$  \\
$\ge 2$ central light jets & $13.2$ & $15.5$  \\ 
$b$ and forward jet veto & $9.5$ & $12.5$  \\ 
$Z$ window & $9.1$ & $11.7$  \\ 
$p_{T,\ell\ell} > 80$~GeV & $8.0$ & $10.2$  \\ 
$E_T^{\rm miss,rel} > 80$~GeV & $5.1$ & $7.0$  \\
$0.3 < \Delta R_{\ell\ell} < 1.5$ & $4.2$ & $5.9$  \\ 
$50 < m_{jj} < 100$~GeV & $2.7$ & $3.6$  \\ 
$p_T(j_1,j_2) > 45$~GeV & $1.8$ & $1.7$  \\ 
\hline 
\end{tabular}
\end{center}
\end{table}

\begin{table}[!t]
\caption{Cut flow for slepton-pair production in SR-$m^{120}_{\rm T2}\,ee$ of ATLAS-SUSY-2013-11, 
for the benchmark point with $(m_{\tilde\ell}, m_{\tilde{\chi}^0_1})=(250,10)$~GeV.}
\label{tab:cutflowMT2120ee_slep25010}
\begin{center}
\begin{tabular}{l|c|c}
cut & ATLAS result & {\sc MA}5 result \\ 
\hline
Initial number of events & & $96.8$  \\ 
2 OS leptons & & $65.3$  \\ 
$m_{\ell\ell} > 20$~GeV & & $65.1$ \\ 
$\tau$ veto & & $65.1$  \\
$ee$ leptons & $51.2$ & $32.1$  \\ 
jet veto & $19.4$ & $17.5$  \\ 
$Z$ veto & $18.7$ & $16.9$  \\ 
$m_{T2} > 120$~GeV & $9.1$ & $8.2$  \\ 
\hline 
\end{tabular}
\end{center}
\end{table}

%%%%%%%%%%%%%%%%%%%%%%%%%%%%%%%%%%%%%%%%%%%%%%%%%%%%%%%%%

\section{Limit setting}\label{sec:CLprocedure}

%%%%%%%%%%%%%%%%%%%%%%%%%%%%%%%%%%%%%%%%%%%%%%%%%%%%%%%%%

For the statistical interpretation of the results, we provide on \cite{ma5wiki} a {\sc Python} code, 
{\tt exclusion\_CLs.py}, for computing exclusions using the $\mathrm{CL}_s$ prescription~\cite{Read:2002hq}.\footnote{The
{\sc Python} code requires {\sc SciPy} libraries to be installed.}
This code can also be installed on a user system by typing in, from the \ma\ interpreter, the command
\begin{verbatim}
  install RecastingTools
\end{verbatim}
which results in the file \texttt{exclusion\_CLs.py} being present at the
root of any working directory created in the expert mode of \ma. 
We refer to~\cite{Conte:2014zja,ma5wiki} for details on the creation of \ma\ working directories.

The \texttt{exclusion\_CLs.py} code takes as input the acceptance $\times$ efficiency information 
from the cut flow {\sc Saf} files generated when executing
an analysis implemented in \ma\ (see Section~\ref{sec:MA5SRM}).
Moreover, an {\sc Xml} file, named {\tt analysis\_name.info} (where \texttt{analysis\_name} stands for a generic
analysis name), needs to be provided by the user in the {\tt Build/Sam\-ple\-Analyzer/U\-ser/A\-na\-ly\-zer} directory, 
specifying the luminosity {\tt <lumi>}, 
the number of observed events {\tt <nobs>}, 
the nominal number of expected SM background events {\tt <nb>}, 
and its uncertainty at 68\%~CL {\tt <deltanb>} in each of the regions, as given in the experimental publication.
The syntax of this file is as follows:

\begin{verbatim}
 <analysis id="cms_sus_13_011">
   <lumi>19.5</lumi> <!-- in fb^-1 -->

   <region type="signal" id="SRname">
     <nobs>227</nobs>
     <nb>251</nb>
     <deltanb>50</deltanb>
   </region>
   ...
   ...
 </analysis>  
\end{verbatim}

\noindent
The attribute {\tt type} of the root tag {\tt <analysis>} can be {\tt signal} or {\tt control} and is optional (the default value is {\tt signal}). 
The {\tt id} of each {\tt <region>} tag has to match the exact name of the SR used in the analysis code. 
When results are given after combining several SRs (for example, for same-flavor leptons instead of $ee$ and $\mu\mu$ separately), the relevant SRs should all be listed  in the attribute {\tt id} separated by semicolons (without extra space). 
Taking the example of the ATLAS analysis presented in Section~\ref{sec:atlas-susy-13-11}, this would read

\begin{verbatim}
 <region id="MT2-90 ee;MT2-90 mumu">
\end{verbatim}

The last piece of information essential for calculating exclusions is the signal cross section.
It can be provided by the user in the {\sc Saf} file {\tt mypoint.txt.saf} (automatically generated when
executing an analysis, see Section~\ref{sec:MA5SRM}), 
where {\tt mypoint.txt}, stored in the \texttt{Input} folder of the working directory, 
is the input file for running the analysis under consideration. 
Alternatively, the cross section can be given as argument when calling {\tt exclusion\_CLs.py}.
Concretely, the limit-setting code is called as
\begin{verbatim}
 ./exclusion_CLs.py analysis_name mypoint.txt \
         [run_number] [cross section in pb]
\end{verbatim}
where the run number and cross section value are optional arguments. 
The run number $x$ (default zero) identifies the output directory to use, 
as each execution of the analysis code yields the creation of a new output directory, 
\texttt{analysis\_name\_}$x$, for the $x^{\rm th}$ execution of the analysis code (starting from 0). 

The procedure of {\tt exclusion\_CLs.py} starts by selecting
the most sensitive SR ({\it i.e.}, the one that yields the best expected exclusion, assuming that
the number of observed events is equal to the nominal number of background events).
This is a standard procedure at the LHC whenever the SRs defined in the analysis are overlapping; 
here we use it as the default for all analyses. 
Then the actual exclusion is calculated, and the confidence level with which the tested
scenario is excluded using the $\mathrm{CL}_s$ prescription~\cite{Read:2002hq} is printed on the screen 
together with the name of the most sensitive SR. The same information is also
stored in the file {\tt a\-na\-ly\-sis\_na\-me\_}$x${\tt .out}, located in the working directory 
of the  \texttt{Output} folder.
Last but not least, if a negative number is given for the cross section, the code returns instead the nominal cross section
that is excluded at 95\% CL, computed using a root-finding algorithm.

The core of the calculation works as follows. First, the number of signal events ($n_s$) is obtained as the
product of the luminosity, signal cross section and acceptance $\times$ efficiency for the SR of interest.
This is used, together with the number of observed events ($n_{\rm obs}$) and the nominal number of background
events ($\hat n_b$) and its uncertainty ($\Delta n_b$) to compute the exclusion.
A large number of toy MC experiments ($10^5$ by default) are then generated from the Poisson distribution
${\rm poiss}(n_{\rm obs} | n_{\rm expected})$, 
corresponding to the distribution of the total number of events in the SR under the
background-only hypothesis on the one hand ($n_{\rm expected} = n_b$), and under the
signal $+$ background hypothesis ($n_{\rm expected} = n_s + n_b$) on the other hand.
We assume that the uncertainty on the number of background events is modeled as ${\rm gauss}(\hat n_b, \Delta n_b)$,
and for each toy MC the number of background events $n_b$ is randomly generated from this normal distribution.
Under the two different hypotheses, $p$-values are then calculated using the number of events actually observed at the LHC, and finally used to compute the $\mathrm{CL}_s$ value.

\begin{figure}[!t]\centering
\vspace*{-2mm}
\includegraphics[width=6cm]{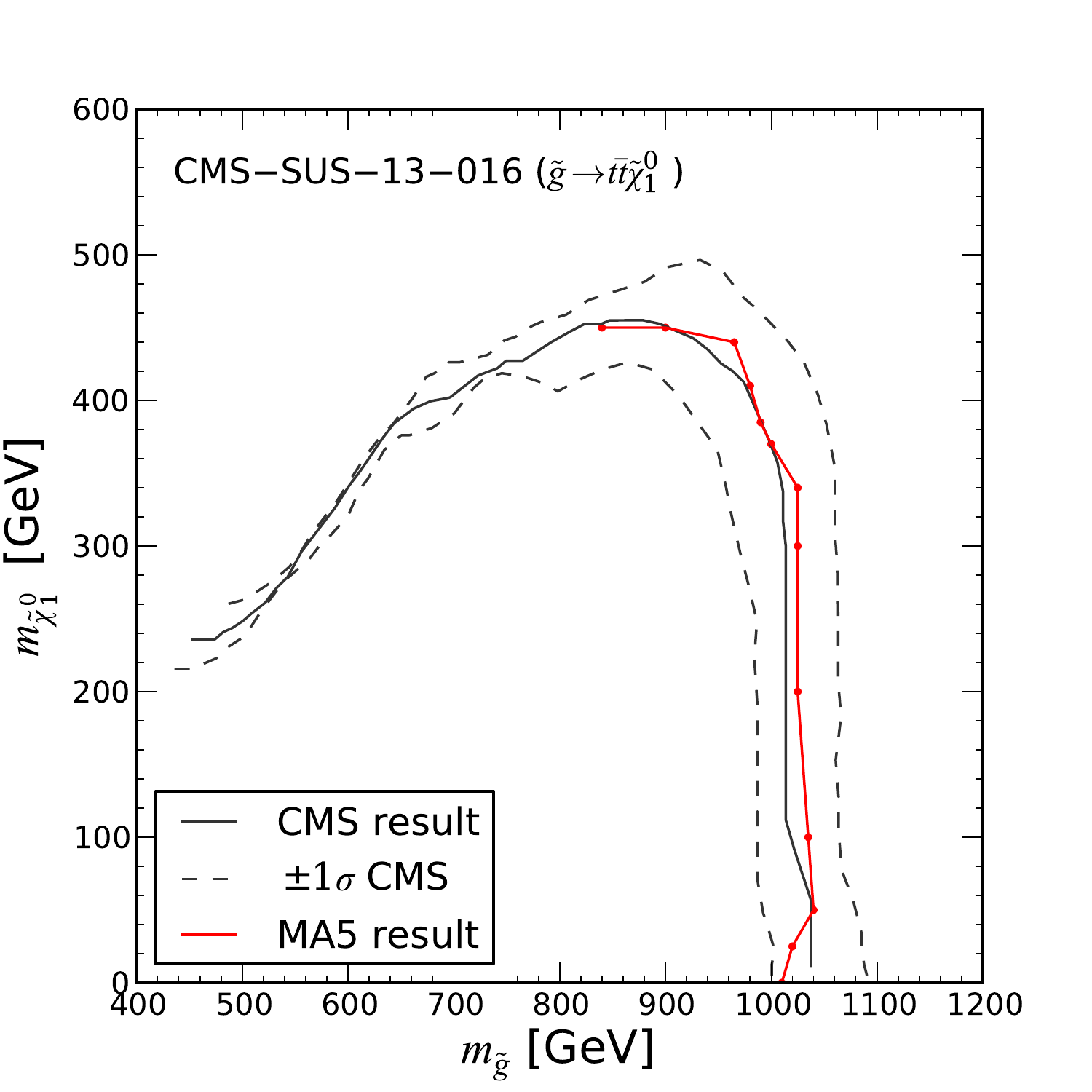}
\caption{The 95\%~CL exclusion limit (in red) in the $\tilde\chi^0_1$ versus $\tilde g$ mass plane reproduced 
with the \ma\ implementation  \cite{ma5code:cms-sus-13-016}  of CMS-SUS-13-016. For comparison, the full and dashed grey lines show the official CMS result with its $\pm1\sigma$ uncertainty from Fig.~6 of \cite{CMS-PAS-SUS-13-016}. 
The limit setting in the region where one of the tops from the gluino decay is off-shell, \ie\ for $m_{\tilde g}\lesssim 800$~GeV, is work in progress.} 
\label{fig:cms-016-limit}
\end{figure}

\begin{figure}[!t]\centering
\vspace*{2.2mm}
\includegraphics[width=5.8cm]{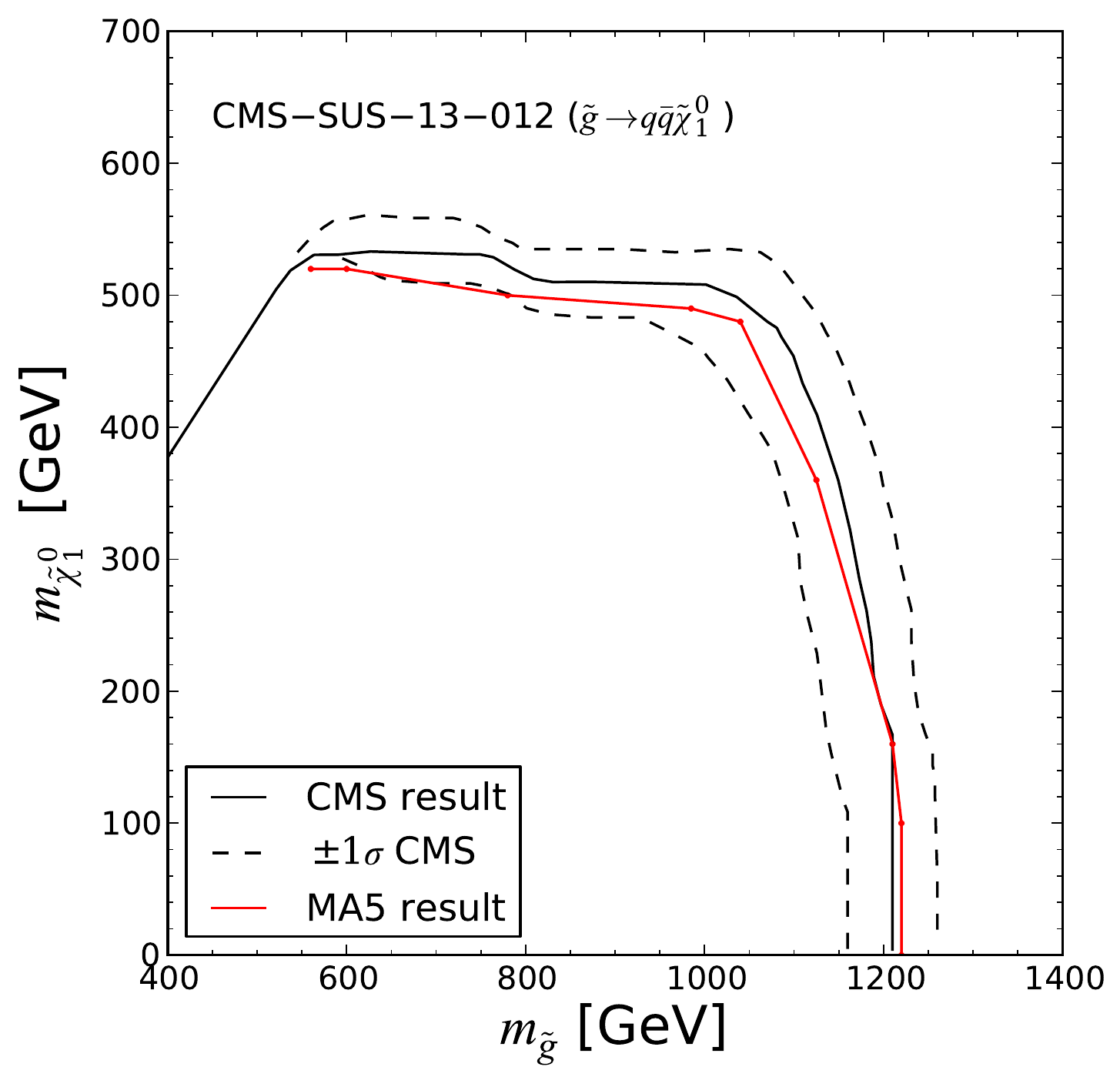}
\caption{The 95\%~CL exclusion limit in the $\tilde\chi^0_1$ versus $\tilde g$ mass plane for the 
$\tilde g\to q\bar q\tilde\chi^0_1$ topology, T1qqqq, reproduced with the \ma\ implementation  \cite{ma5code:cms-sus-13-012} of CMS-SUS-13-012.}
\label{fig:cms-012-limit} 
\end{figure}

\begin{figure}[!t]\centering
\vspace*{2.2mm}
\includegraphics[width=5.8cm]{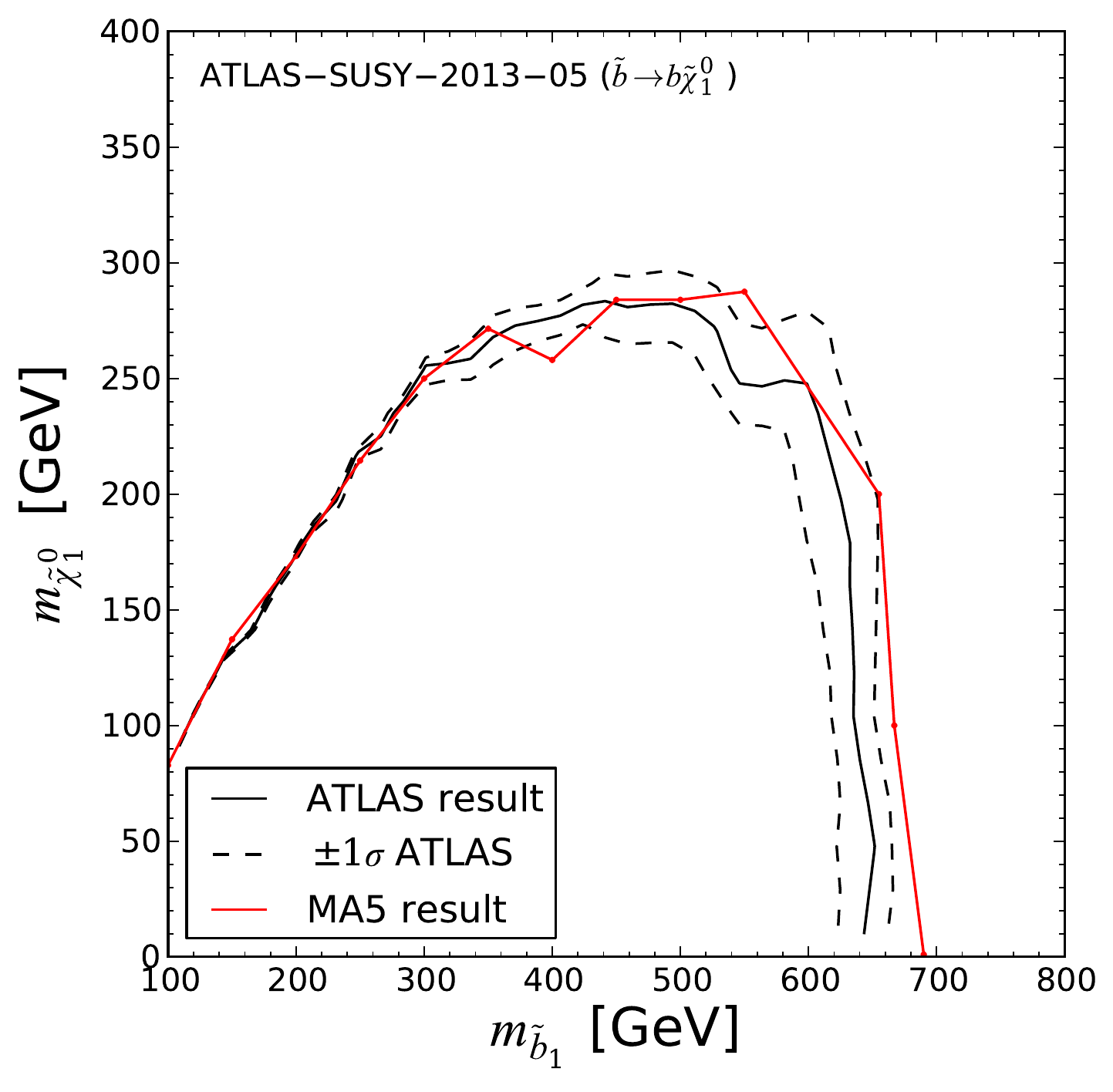}
\caption{The 95\%~CL exclusion limit in the $\tilde\chi^0_1$ versus $\tilde b_1$ mass plane for the 
$\tilde b_1\to b\tilde\chi^0_1$ topology, reproduced with the \ma\ implementation~\cite{ma5code:atlas-susy-2013-05}  
of ATLAS-SUSY-2013-05.}
\label{fig:atlas-05-limit} 
\end{figure}

We have tested the limit-setting code on the analyses presented in this paper and generally found good agreement with the official exclusions from ATLAS and CMS. Figures~\ref{fig:cms-016-limit}--\ref{fig:atlas-05-limit} give some 
illustrative examples. % (more can be found in the validation notes). 
In particular, 
Fig.~\ref{fig:cms-016-limit} shows the 95\%~CL exclusion limit in the neutralino versus gluino mass plane 
for the $\tilde g\to t\bar t\tilde\chi^0_1$ topology 
reproduced with the \ma\ implementation \cite{ma5code:cms-sus-13-016}  of CMS-SUS-13-016.  
This analysis has only one SR and thus provides a good test for our implementation of the $\mathrm{CL}_s$ prescription. 
To prove that our procedure also works well for analyses with many SRs, 
Fig.~\ref{fig:cms-012-limit} shows the 95\%~CL exclusion limit for the $\tilde g\to q\bar q\tilde\chi^0_1$ topology (T1qqqq)
reproduced with the \ma\ implementation \cite{ma5code:cms-sus-13-012}  of CMS-SUS-13-012.  
We also find good agreement for the T1tttt and T5VV topologies for this analysis; the one case that works less well is the T2qq topology (squark-pair production with $\tilde q\to q\tilde\chi^0_1$) for which the reproduced limit becomes unreliable for neutralino masses above about 200--250~GeV. For improving the situation, the statistical model for combining the SRs   %and treating background correlations 
would be needed from CMS, but this is not available. 
Finally, Fig.~\ref{fig:atlas-05-limit} shows the 95\%~CL exclusion limit in the neutralino versus sbottom mass plane for 
$\tilde{b}_1$ pair production with $\tilde{b}_1 \rightarrow b\tilde{\chi}_1^0$ reproduced with the \ma\ implementation~\cite{ma5code:atlas-susy-2013-05} of ATLAS-SUSY-2013-05 and compared to the official ATLAS limit. This is the case where the largest differences are encountered in the kinematic distributions, see Section~\ref{sec:atlas-susy-13-05}. None the less we see that the limit is reasonably well reproduced (note that the $\pm 1\sigma$ uncertainty quoted by ATLAS is based only on the theory uncertainty of the cross section).

%%% new %%%
%We note, however, 
Last but not least it is important to note that the module 
{\tt exclusion\_CLs.py} is intended only as a lightweight tool for the user who wants an 
approximate but fast evaluation of the results of his/her simulation.  Users who want to go beyond the simplifications made in {\tt exclusion\_CLs.py} are encouraged to use \eg\ the {\tt RooFit} and {\tt RooStats} machinery \cite{rooCLs} 
adopted by ATLAS and CMS.  
%%%

%%%%%%%%%%%%%%%%%%%%%%%%%%%%%%%%%%%%%%%%%%%%%%%%%%%%%%%%%

\section{Guidelines}\label{sec:guidelines}

%%%%%%%%%%%%%%%%%%%%%%%%%%%%%%%%%%%%%%%%%%%%%%%%%%%%%%%%%

In this section, we provide some brief guidelines, on the one hand for the experimental collaborations regarding what material is needed for a reliable implementation and validation of an analysis, on the other hand for potential contributors to the framework as to how to validate a new analysis implementation.  

\subsection{Information needed from the experiments}

The information needed from the experimental collaboration on an analysis splits into two parts: analysis description and material for validation. We ask the collaborations to provide\\

\noindent 1.~in the analysis description:
\begin{itemize}
\item a clear and unambiguous definition of all the cuts and the sequence in which they are applied;
\item efficiencies as function of $p_T$ (and, where relevant, $\eta$) for all physics objects considered
in the analysis: electrons, muons, taus, $b$-jets, light jets, {\it{etc.}};
\item efficiencies for triggers and event cleaning.
\end{itemize}
At present, cuts are typically well defined, but their sequence is not always clear. A clear ordering in tabulated form or by means of detailed cut flows would help. Efficiencies are sometimes only roughly indicated, which is very problematic for us; if an efficiency is not given explicitly in the paper, it should be clearly referenced where to find it (\eg\ pointing to the precise figure in a performance note). 
Efficiencies for triggers and event cleaning are of particular concern, as they are highly important for our purpose but currently often missing altogether in the experimental papers.\\ 

\noindent 2.~as validation material:
\begin{itemize}
\item unambiguously defined benchmark points, \eg\ in the form of SLHA files (including
the full mass spectrum and decay tables) and/or parton-level MC event files;
\item exact configuration of MC tools:  the ideal would be if the run cards and input scripts for {\sc MadGraph}, {\sc Pythia}, {\it etc.}\ were made available; if this is not the case, we need at least the exact versions of the MC tools and their basic settings;
\item detailed cut flows for all benchmark points, showing each step of the analysis;
\item histograms of kinematic distributions after specific cuts.
\end{itemize}

\noindent 
Only if complete information is provided for an analysis can the recasting be done in a reliable way. 
In this respect is should also be noted that for any analysis in which SRs are combined %for the interpretation of the results, 
the corresponding likelihood model should be made available by the collaboration. 

Some more comments are in order. First, we note that code modules for special kinematical variables, as currently provided by CMS, are extremely useful. We highly appreciate this practice.  
Second, we note that having to read efficiencies, event numbers or other data off paper plots is very tedious and introduces unwarranted uncertainties, especially when dealing with log-scale plots.  We therefore strongly encourage the collaborations to always provide their plots also in numerical form, be it on {\sc HepData} or on the analysis Twiki page.
Finally, one could also imagine that the experimental collaborations directly provide validated \madanalysis\ implementations for certain analyses. While this would be an excellent way of documenting an analysis, 
this is of course left to the initiative and decision of the respective search groups.

\subsection{Recommendations for implementing and validating new analyses}

Since the framework we presented here is intended as an open-source project, we also give some guidelines for potential contributors: 

\begin{itemize}
\item clearly identify and reference the analysis together with your contact details in the header of the recast code; 
\item always implement all SRs of an analysis;
\item take care that the code is clean and well commented;
\item reproduce all the cut flows provided by the experimental collaboration for the various benchmark points;
\item reproduce all the available kinematic distributions for  the benchmark points;
\item for the above, use the exact same settings of the MC tools as the experimental collaboration;
\item if information for any of the above is missing, contact the experimental collaboration;
\item likewise, contact the experimental collaboration if not enough validation material is available, \eg\ if cut flows are not detailed enough;
\item the required agreement with the experimental results is somewhat analysis dependent --- generally, we think it should be of the order of 30\% or better for the final numbers as well as in each step of the cut flow;  if larger discrepancies are found, contacting the collaboration can be helpful for resolving them; 
\item while we think that cut flows and kinematical distributions should be the primary validation material, it is also a good idea to reproduce the 95\% CL limit curve for the relevant simplified model(s); 
\item publish your code via {\sc Inspire}~\cite{inspiresubmission};
\item provide a detailed validation note to be put on~\cite{ma5wiki}.
\end{itemize}

%\clearpage
%%%%%%%%%%%%%%%%%%%%%%%%%%%%%%%%%%%%%%%%%%%%%%%%%%%%%%%%%

\section{Conclusions}\label{sec:conclusions}

%%%%%%%%%%%%%%%%%%%%%%%%%%%%%%%%%%%%%%%%%%%%%%%%%%%%%%%%%

We have presented a new scheme for developing and deploying implementations of LHC analyses 
based on fast simulation within the \ma\ framework. This can serve to create a public analysis database,  
which may be used and developed further by the whole community.  
The codes for the five analyses~\cite{ma5code:cms-sus-13-011,ma5code:cms-sus-13-012,ma5code:cms-sus-13-016,ma5code:atlas-susy-2013-05,ma5code:atlas-susy-2013-11} 
that we published together with this paper are intended as a starting point 
for this database and may conveniently be used as templates for other analyses.  

We propose that the C++ codes of new implementations within this scheme be published 
via {\sc Inspire}~\cite{inspiresubmission}, as done here,  
best together with the physics paper they have been developed for. 
This way, each analysis implementation is assigned a 
Digital Object Identifier (DOI)~\cite{doi}, ensuring that it is uniquely identifiable, searchable and citable. 
In addition it is very useful if a detailed validation note is made available on the  \ma\ wiki page~\cite{ma5wiki}. 

The ease with which an experimental analysis can be implemented and validated may serve as a useful check for the experimental collaborations for the quality of their documentation. Note, finally, that the platform we are pro\-posing might also be used by the experimental collaborations to directly provide implementations of their analyses for fast simulation, thereby assuring the maximum usability of their results, as for example envisaged in level 1 of the CMS statement on ``data preservation, re-use and open access policy''~\cite{CMS:DataPolicy}. 

It is important for the legacy of the LHC that its experimental results can be used by the whole high-energy physics community. We hope that our project contributes to this aim.

%%%%%%%%%%%%%%%%%%%%%%%%%%%%%%%%%%%%%%%%%%%%%%%%%%%%%%%%%

\section*{Acknowledgements}

We thank our experimental colleagues, in particular 
Aless\-an\-dro Gaz, Benjamin Hooberman, Christian Sander, Niki Saoulidou, Nadia Strobbe, 
Keith A.\ Ulmer, Frank W\"urth\-wein from CMS, as well as  
Jamie Boyd, Marie-Helene Genest and Monica d'Onofrio from ATLAS 
for help in the validation process and for providing additional information where needed. 
Moreover, we thank Daniel Schmeier and Jamie Tattersall for extremely useful discussions. 
This work originated from the 2013 Les Houches ``Physics at TeV Colliders'' workshop; 
the inspiring and open-minded atmosphere of this workshop has been crucial for this project.  

Partial financial support by the French ANR projects 
12-BS05-0006 DMAstroLHC  and 12-JS05-002-01 BATS@ LHC, 
the Theory-LHC-France initiative of CNRS/IN2P3, 
and the ``Investissements d'avenir, Labex ENIGMASS'' is gratefully acknowledged. 
S.B.~thanks the LPSC Grenoble for hospitality.

%%%%%%%%%%%%%%%%%%%%%%%%%%%%%%%%%%%%%%%%%%%%%%%%%%%%%%%%%
%\bibliographystyle{utphys}
%\bibliography{references}
%%%%%%%%%%%%%%%%%%%%%%%%%%%%%%%%%%%%%%%%%%%%%%%%%%%%%%%%%

\providecommand{\href}[2]{#2}\begingroup\raggedright\endgroup

%%%%%%%%%%%%%%%%%%%%%%%%%%%%%%%%%%%%%%%%%%%%%%%%%%%%%%%%%

\end{document}